\documentclass[a4paper,11pt]{article}
\pdfoutput=1
\usepackage{jheppub}
\usepackage{color}
\usepackage{array}
\newcolumntype{C}[1]{>{\centering\arraybackslash}p{#1}}
\usepackage{slashed,dsfont}
\usepackage{bm,wasysym}
\usepackage{longtable}
\usepackage{booktabs}

\usepackage{tabularx}
\usepackage{rotating}
\usepackage{lscape}

\newcommand{\lsim}{
\mathrel{\hbox{\rlap{\hbox{\lower4pt\hbox{$\sim$}}}\hbox{$<$}}}}

\newcommand{\gsim}{
\mathrel{\hbox{\rlap{\hbox{\lower4pt\hbox{$\sim$}}}\hbox{$>$}}}}

\renewcommand{\arraystretch}{2}

\allowdisplaybreaks[2]

\newcommand{\nn}{\nonumber}

\def\mLb{{m_{\Lambda_b}}}
\def\mmLb{{m^2_{\Lambda_b}}}
\def\mL{{m_{\Lambda}}}
\def\mmL{{m^2_{\Lambda}}}

\def\plpl{{+\frac{1}{2}+\frac{1}{2}}}
\def\plmi{{+\frac{1}{2}-\frac{1}{2}}}
\def\mipl{{-\frac{1}{2}+\frac{1}{2}}}
\def\mimi{{-\frac{1}{2}-\frac{1}{2}}}
\def\re{{\rm Re}}  \def\im{{\rm Im}}
\def\mC{{\mathcal{C}}}
\def\mK{{\mathcal{K}}}

\definecolor{schrift}{RGB}{120,0,0}

\def\ARpe#1{{A^R_{\perp_{#1}}}}  \def\ARpa#1{{A^R_{\|_{#1}}}}

\def\AsRpe#1{{A^{\ast R}_{\perp_{#1}}}}  \def\AsRpa#1{{A^{\ast R}_{\|_{#1}}}}

\title{\boldmath\color{schrift}{On the angular distribution of $\Lambda_b\to\Lambda(\to N\pi)\tau^+\tau^-$ decay}}

\author{Diganta Das}

\affiliation{{\sf Department of Physics and Astrophysics, University of Delhi, Delhi 110007, India}}

\emailAdd{diganta99@gmail.com}

\abstract{We present a full angular distribution of the four body $\Lambda_b\to\Lambda(\to N\pi)\ell^+\ell^-$ decay where the leptons are massive and the $\Lambda_b$ is unpolarized, in an operator basis which includes the Standard Model operators, new vector and axial-vector operators, and scalar and pseudo-scalar operators. The angular coefficients are expressed in terms of transversity amplitudes. We study several $\Lambda_b\to\Lambda(\to p\pi)\tau^+\tau^-$ observables in the Standard Model and in the presence of the new operators. For our numerical analysis, we use the form factors from lattice QCD calculations. }

\keywords{Rare Decays, Baryon Decays, New Physics}

\begin{document}

\maketitle

\renewcommand{\arraystretch}{1.6}

\section{Introduction}
The loop induced rare $b\to s\ell^+\ell^-$ transitions are well known for their high sensitivity to physics beyond the Standard Model (SM), also known as new physics (NP). The $b\to s\ell^+\ell^-$ mediated decay of $B$ meson, the $B\to K^\ast(\to K\pi) \ell^+\ell^-$, is a very important decay mode in this regard as its full angular distribution gives access to a multitude of observables \cite{Kruger:1999xa} that can uniquely test the SM. Therefore, the $B\to K^\ast(\to K\pi) \ell^+\ell^-$ decay is the topic of intense theoretical and experimental scrutiny in the past several decades.  Recently, the LHCb measured \cite{Aaij:2015xza} the first $b\to s\mu^+\mu^-$ mediated baryonic decay $\Lambda_b\to\Lambda\mu^+\mu^-$ which was earlier observed by the CDF \cite{Aaltonen:2011qs}. The full angular distribution of $\Lambda_b\to\Lambda(\to N\pi)\mu^+\mu^-$ ($N\pi\equiv\{p^+\pi^-, n^0\pi^0\}$) with unpolarized $\Lambda_b$ also provides a large number of observables \cite{Boer:2014kda} which makes it at par with the $B\to K^\ast(\to K\pi) \ell^+\ell^-$ as far the number of accessible observables is concerned. On the other hand, if the polarization of the $\Lambda_b$ is accounted for, the number of observables that one can gain access to become more than three times than the unpolarized case \cite{Blake:2017une} making this decay an unique test ground for the SM and searches for NP.

While the short-distance physics of $\Lambda_b\to\Lambda(\to N\pi)\ell^+\ell^-$, encoded in the $b\to s\ell^+\ell^-$ Wilson coefficients are very precisely known for some time \cite{Bobeth:1999mk}, progress in the long distance calculations has recently been made. This includes the QCD sum-rules analysis of spectator-scattering corrections to the form factor relations \cite{Feldmann:2011xf}, and better theoretical understanding of light-cone distribution amplitudes of $\Lambda_b$ baryon \cite{Ali:2012pn,Bell:2013tfa,Braun:2014npa}. At low dilepton invariant mass squared $q^2$ or large recoil, the form factors are calculated in the light-cone sum rules in Refs.~\cite{Wang:2008sm,Wang:2015ndk}, and in soft-collinear effective theory in Refs.~\cite{Wang:2011uv,Mannel:2011xg}. The form factors have also been calculated in the relativistic quark model in Ref.~\cite{Faustov:2017wbh}. For our need of the form factors at large $q^2$ or low recoil, we use the results from the calculations in lattice QCD \cite{Detmold:2016pkz,Detmold:2012vy}

In a recent paper \cite{Das:2018sms} we presented a model independent NP analysis of $\Lambda_b\to\Lambda\ell^+\ell^-$ decay where the NP operators include new vector and axial vector (VA) operators, scalar  and pseudo-scalar (SP) operators, and tensor operators. In this analysis the mass of the final state leptons was neglected considering di-muon in the final state. The lepton mass should however be incorporated if there are $\tau^+\tau^-$ in the final state, or one wants to study lepton flavor universality (LFU) violation. Recent SM analysis of $\Lambda_b\to\Lambda(\to N\pi)\ell^+\ell^-$ with massive leptons can be found in Refs.~\cite{Roy:2017dum,Blake:2017une,Gutsche:2013pp}. In this paper we present a full angular analysis of $\Lambda_b\to\Lambda(\to N\pi)\ell^+\ell^-$ in the SM+NP operator basis where we consider the leptons to be massive and the $\Lambda_b$ to be unpolarized. The NP operators that we consider are new VA operators and SP operators. To keep the analysis simple, we have ignored the tensor operators as these operators lead to a large number of terms in the angular coefficients. We work in the transversity basis and calculate the angular coefficients in terms of fourteen transversity amplitudes. The full angular distribution allows us to construct a number of observables that can be measured in experiments. For phenomenological analysis, we study the $\Lambda_b\to\Lambda(\to p\pi)\tau^+\tau^-$ decay and present determinations  of several observables in the SM and in the presence of the NP couplings. At present, no $b\to s\tau^+\tau^-$ transition has yet been observed and there are only upper limits on $B_s\to\tau^+\tau^-$ and $B^+\to K^+\tau^+\tau^-$ from the LHCb \cite{Aaij:2017xqt} and BaBar \cite{TheBaBar:2016xwe}, respectively. Interestingly, it has been shown in Refs.~\cite{Alonso:2015sja,Crivellin:2017zlb,Capdevila:2017iqn} that an explanation of the flavor anomalies in the recent $b\to c\ell\nu$ and $b\to s\mu^+\mu^-$ transitions \cite{Aaij:2014ora, Aaij:2017vbb, Aaij:2017tyk, HFAG2017} lead to large enhancement in $b\to s\tau^+\tau^-$ rates. The $\Lambda_b\to\Lambda(\to p\pi)\tau^+\tau^-$ mode is therefore worth exploring.

The paper is organized as follows. In Sec.~\ref{subsec:effHam} we give the effective Hamiltonian for $b\to s\ell^+\ell^-$ transition. In Sec.~\ref{sec:kin} we describe the decay kinematics and work out the expressions of the hadronic and leptonic amplitudes in Sec.~\ref{sec:amplitudes}. The full angular distribution of $\Lambda_b\to\Lambda(\to N\pi)\ell^+\ell^-$ is derived in Sec.~\ref{sec:angdist}. The $\Lambda_b\to\Lambda$ form factors are described in Sec.~\ref{sec:ff} and we perform the numerical analysis in Sec.~\ref{sec:numerical}. The results are summarized in Sec.~\ref{sec:summary}. We also give several appendixes where details of the derivations can be found.

\section{Effective Hamiltonian \label{subsec:effHam}}
In the SM the rare $b \to s\ell^+\ell^-$ transition proceeds through loop diagrams and contains the radiative operator $\mathcal{O}_7$, the semi-leptoic operators $\mathcal{O}_{9,10}$, and the hadronic operators $\mathcal{O}_{1-6,8}$. We assume only the factorizable quark loop corrections to the hadronic operators which can be absorbed into the Wilson coefficients $\mC_{7,9}^{\rm eff}$ corresponding to the operators $\mathcal{O}_{7,9}$. For simplicity we ignore the non-factorizable corrections which are expected to play significant role, particularly at large recoil \cite{Beneke:2001at,Beneke:2004dp}. Going beyond the SM operator basis, we also include new VA operators and SP operators. Therefore our Lagrangian reads \cite{Das:2018sms}
\begin{equation}\label{eq:Heff1}
\mathcal{H}^{\rm eff} = - \frac{4G_F}{\sqrt{2}}V_{tb}V_{ts}^\ast\frac{\alpha_e}{4\pi}  \bigg(\sum_i \mC_i \mathcal{O}_i + \sum_j  \mC^\prime_j \mathcal{O}^\prime_j \bigg)\, ,
\end{equation}
where $i = 7, 9, 10, V, A, S, P$ and $j = V, A, S, P$. The operators $\mathcal{O}^{(\prime)}$ read 
\begin{eqnarray}\label{eq:opbasis}
\begin{split}
&O_7^{} = \frac{m_b}{e} \big[\bar{s}\sigma^{\mu\nu}P_{R}b\big]F_{\mu\nu}\, ,\quad \mathcal{O}_9 = \big[\bar{s}\gamma^\mu P_{L}b \big]\big[\ell\gamma_\mu\ell \big]\, ,
\quad \mathcal{O}_{10} = \big[\bar{s}\gamma^\mu P_{L}b \big]\big[\ell\gamma_\mu\gamma_5\ell \big]\, ,\\
&\mathcal{O}^{(\prime)}_V = \big[\bar{s}\gamma^\mu P_{L(R)}b \big]\big[\ell\gamma_\mu\ell \big]\, ,
\quad \mathcal{O}^{(\prime)}_{A} = \big[\bar{s}\gamma^\mu P_{L(R)}b \big]\big[\ell\gamma_\mu\gamma_5\ell \big]\, ,\\
&\mathcal{O}_S^{(\prime)} = \big[\bar{s}P_R(L)b \big]\big[\ell\ell \big]\, ,\quad \mathcal{O}_P^{(\prime)} = \big[\bar{s}P_R(L)b \big]\big[\ell\gamma_5\ell \big]\, .
\end{split}
\end{eqnarray}
The radiative or dipole operator $\mathcal{O}_7$ contributes to the $b\to s\ell^+\ell^-$ transition through its coupling to a dilepton pair 
\begin{equation}
 \langle \Lambda(k)\ell^+(q_1)\ell^-(q_2)|\mathcal{O}_7|\Lambda_b(p) \rangle = -\frac{2m_b}{q^2} \langle \Lambda |  \bar{s}i\sigma^{\mu\nu} q_\nu P_{R}b\big | \Lambda_b \rangle (\bar{u}_\ell \gamma_\mu v_\ell)\, .
\end{equation}
Here $G_F$ denotes the Fermi-constant, $\alpha_e=e^2/4\pi$ is the fine structure constant, $V_{tb}V_{ts}^\ast$ are the Cabibbo-Kobayashi-Maskawa(CKM) elements, $P_{L,R}=(1\mp\gamma_5)/2$ are the chiral projection operators, and $\sigma_{\mu\nu}=i[\gamma_\mu,\gamma_\nu]$/2. The $b$-quark mass appearing in the dipole operator $\mathcal{O}_7$ is the running mass in the modified minimal subtraction ($\overline{\text{MS}}$) scheme. In the SM, the Wilson coefficients of the NP operators $\mC_{V,A}^{(\prime)},\mC_{S,P}^{(\prime)}$ vanish. The expressions of $\mC_{9}^{\rm eff}$ and other SM Wilson coefficients are given in Appendix \ref{app:C910}. In Eq.~(\ref{eq:Heff1}) we have neglected additional Cabibbo-suppressed contributions proportional to $V_{ub}V_{us}^\ast$ since $V_{ub}V_{us}^\ast<<V_{tb}V_{ts}^\ast$. 

\section{Kinematics of $\Lambda_b\to\Lambda(\to N\pi)\ell^+\ell^-$ decay \label{sec:kin}}
We assign the following momenta and spin variables to the different particles in the decay process 
\begin{eqnarray}
\begin{split}
&\Lambda_b(p,s_p ) \to \Lambda(k,s_k) \ell^+(q_1) \ell^-(q_2)\, ,\\
&\Lambda(k,s_k) \to N(k_1,s_N)\pi(k_2)\, ,
\end{split}
\end{eqnarray}
\emph{i.e.,} $p,k,k_1,k_2,q_1$ and $q_2$ are the momenta of $\Lambda_b$, $\Lambda$, $N$, $\pi$, and the positively and negatively charged leptons, respectively, and $s_{p,k,N}$ are the projections of the baryonic spins on to the $z$-axis in their respective rest frames. From momentum conservation $k = k_1+k_2$ and we defined the four momentum of the dilepton pair as
\begin{equation}
q^\mu = q_1^\mu + q_2^\mu\, .
\end{equation}

The decay can be completely described in terms of four independent kinematic variables which we choose as the dilepton invariant mass squared $q^2$, the angle $\theta_\ell$ which is defined as the angle made by $\ell^-$ with the $+z$-direction in the dilepton rest frame, the angle $\theta_\Lambda$ which is made by $N$ with respect to the $+z$-direction in the $N\pi$ rest frame, and the angle $\phi$ between the $\ell^+\ell^-$ and $N\pi$ decay planes. 

\section{Decay amplitudes \label{sec:amplitudes}} 
The four body decay proceeds in two steps, the decay $\Lambda_b\to \Lambda\ell^+\ell^-	$ followed by $\Lambda\to N\pi$. Corresponding to the Hamiltonian (\ref{eq:Heff1}) the amplitudes for $\Lambda_b\to \Lambda\ell^+\ell^-$ can be written as \cite{Das:2018sms}
\begin{eqnarray}\label{eq:Mll}
	\mathcal{M}^{\lambda_1,\lambda_2}(s_p,s_k) &=& - \frac{G_F}{\sqrt{2}}V_{tb}V_{ts}^\ast \frac{\alpha_e}{4\pi} \sum_{i=L,R}\bigg[ \sum_{\lambda} \eta_\lambda H^{i,s_p,s_k}_{\rm VA, \lambda} L^{\lambda_1,\lambda_2}_{i,\lambda} + H^{i,s_p,s_k}_{\rm SP} L^{\lambda_1,\lambda_2}_i  \bigg]\, .
\end{eqnarray}
Here we have separated the left (L) and right (R) handed chiralities of the lepton currents. The $\Lambda_b\to \Lambda$ hadronic helicity amplitudes $H^{i,s_p,s_k}$ are defined as the projections of the $\Lambda_b\to \Lambda$ hadronic matrix elements on the direction of polarization of the virtual gauge boson that decays into dilepton pair. The polarization states of the gauge boson are denoted by $\lambda=t,\pm 1, 0$, the helicities of the leptons are denoted by $\lambda_{1,2}$ and $\eta_t = 1, \eta_{\pm 1, 0}=-1$. $H^{i,s_p,s_k}$ were explicitly derived in Ref.~\cite{Das:2018sms} to which we refer but the expressions are also collected in Appendix \ref{sec:TAs2} for completeness. The leptonic helicity amplitudes $L^{\lambda_1,\lambda_2}$ for massive leptons are worked out in Sec.~\ref{sec:LepHel}. 
 
\subsection{Transversity amplitudes \label{sec:TAs}}
In this section we introduce the transversity amplitudes which are defined as linear combinations (see Appendix \ref{sec:TAs2}) of the helicity amplitudes\footnote{In our definition of the transversity amplitudes, the $\perp_1$ and $\perp_0$ ($\|_1$ and $\|_0$) components depend on vector (axial-vector) current. In this regard our SM amplitudes are identical to that in Ref.~\cite{Boer:2014kda}. A different definition is used in Ref.~\cite{Blake:2017une,Gutsche:2013pp} where $\perp_1$ and $\perp_0$ ($\|_1$ and $\|_0$) depend on the axial-vector (vector) current.}. As shown in Eq.~(\ref{eq:Mll}), the amplitudes corresponding to the VA and SP operators can be separated. For VA operators we obtain ten transversity amplitudes
\begin{eqnarray}
A^{L,(R)}_{\perp_1} &=& -\sqrt{2}N \bigg( f^V_\perp \sqrt{2s_-} \mC^{L,(R)}_{\rm VA+} + \frac{2m_b}{q^2} f^T_\perp (\mLb + \mL) \sqrt{2s_-} \mC_7^{\rm eff} \bigg)\, ,\\
A^{L,(R)}_{\|_1} &=& \sqrt{2}N \bigg( f^A_\perp \sqrt{2s_+} \mC^{L,(R)}_{\rm VA-} + \frac{2m_b}{q^2} f^{T5}_\perp (\mLb - \mL) \sqrt{2s_+} \mC_7^{\rm eff} \bigg)\, ,\\
A^{L,(R)}_{\perp_0} &=& \sqrt{2}N \bigg( f^V_0 (\mLb + \mL) \sqrt{\frac{s_-}{q^2}} \mC^{L,(R)}_{\rm VA+} + \frac{2m_b}{q^2} f^T_0\sqrt{q^2s_-} \mC_7^{\rm eff} \bigg)\, ,\\
A^{L,(R)}_{\|_0} &=& -\sqrt{2}N \bigg( f^A_0 (\mLb - \mL) \sqrt{\frac{s_+}{q^2}} \mC^{L,(R)}_{\rm VA-} + \frac{2m_b}{q^2} f^{T5}_0\sqrt{q^2s_+} \mC_7^{\rm eff} \bigg)\, ,\\
A_{\perp t} &=& -2\sqrt{2}N f^V_t (\mLb - \mL) \sqrt{\frac{s_+}{q^2}} (\mC_{10} + \mC_A + \mC_A^\prime)\, ,\\
A_{\| t} &=& 2\sqrt{2}N f^A_t (\mLb + \mL) \sqrt{\frac{s_-}{q^2}} (\mC_{10} + \mC_A - \mC_A^\prime) \, .
\end{eqnarray}

The Wilson coefficients $\mC^{\rm eff}_9,\mC_{10},\mC_{V,A}^{(\prime)}$ appear in the VA amplitudes in the following combinations
\begin{align}
& \mC_{\rm VA,+}^{L(R)} = (\mC_9^{\rm eff}\mp \mC_{10})+(\mC_V\mp \mC_A)+(\mC_V^\prime \mp \mC_A^\prime)\, ,\\
& \mC_{\rm VA,-}^{L(R)} = (\mC_9^{\rm eff}\mp \mC_{10})+(\mC_V\mp \mC_A)-(\mC_V^\prime \mp \mC_A^\prime)\, ,
\end{align}
and we have defined $s_\pm$ as
\begin{equation}\label{eq:spm}
s_\pm = (\mLb \pm \mL)^2 - q^2\, .
\end{equation}
The $q^2$ dependent normalization constant $N$ is given by 
\begin{equation}
N(q^2) = G_F V_{tb}V_{ts}^\ast \alpha_e \sqrt{\tau_{\Lambda_b} \frac{q^2\sqrt{\lambda(\mmLb,\mmL,q^2)}}{3.2^{11} m^3_{\Lambda_b} \pi^5 }\beta_\ell}\, ,\quad \beta_\ell = \sqrt{1 - \frac{4m_\ell^2}{q^2}}\, .
\end{equation}
The current conservation ensures that the timelike amplitudes $A_{\perp t,\|t}$ depend only on the axial-vector couplings. It also ensures that the timelike amplitudes do not contribute if the leptons are massless.

Corresponding to the SP operators we obtain four transversity amplitudes 
\begin{eqnarray}
A_{\rm S\perp} &=& 2\sqrt{2}N f^V_t \frac{\mLb - \mL}{m_b} \sqrt{s_+} (\mC_S + \mC_S^\prime)\, ,\\
A_{\rm S\|} &=& -2\sqrt{2}N f^A_t \frac{\mLb + \mL}{m_b} \sqrt{s_-} (\mC_S - \mC_S^\prime)\, ,\\
A_{\rm P\perp} &=& -2\sqrt{2}N f^V_t \frac{\mLb - \mL}{m_b} \sqrt{s_+} (\mC_P + \mC_P^\prime)\, ,\\
A_{\rm P\|} &=& 2\sqrt{2}N f^A_t \frac{\mLb + \mL}{m_b} \sqrt{s_-} (\mC_P - \mC_P^\prime)\, .
\end{eqnarray}
Note that these amplitudes are proportional to either the scalar ($\mC_S,\mC_S^\prime$) or the pseudo-scalar couplings ($\mC_P,\mC_P^\prime$) only.

 \subsection{Leptonic helicity amplitudes \label{sec:LepHel}}
 The leptonic helicity amplitudes $L^{\lambda_1,\lambda_2}$ are define as
 \begin{align}\label{eq:Ldef1}
 & L^{\lambda_1,\lambda_2}_{L(R)} = \langle \bar{\ell}(\lambda_1)\ell(\lambda_2) | \bar{\ell} (1\mp\gamma_5) \ell | 0\rangle\, , \\
 \label{eq:Ldef2}
 & L^{\lambda_1,\lambda_2}_{L(R),\lambda} = \bar{\epsilon}^\mu(\lambda) \langle \bar{\ell}(\lambda_1) \ell(\lambda_2) | \bar{\ell} \gamma_\mu (1\mp\gamma_5) \ell | 0\rangle\, ,
 \end{align}
 where $\bar{\epsilon}^\mu(\lambda)$ are the polarization vectors of the virtual gauge boson. In Ref.~\cite{Das:2018sms} we derived the expressions of $L^{\lambda_1,\lambda_2}_{L(R)}$ and $L^{\lambda_1,\lambda_2}_{L(R),\lambda}$ in the limit of $m_\ell=0$. When $m_\ell\neq 0$, following the steps in Ref.~\cite{Das:2018sms} we obtain the following non-zero expressions of the leptonic helicity amplitudes 
 for different combinations of $\lambda_1$ and $\lambda_2$ 
 \begin{align}
 & L^{\plpl}_L = -L^{\mimi}_R = \sqrt{q^2}(1+\beta_\ell)\, ,\quad L^{\mimi}_L = -L^{\plpl}_R = \sqrt{q^2}(1-\beta_\ell)\, ,\\
& L^{\plpl}_{L,+1}=L^{\plpl}_{R,+1}=L^{\mimi}_{L,-1}=L^{\mimi}_{R,-1} = \sqrt{2}m_\ell\sin\theta_\ell\, ,\\
& L^{\mimi}_{L,+1}=L^{\mimi}_{R,+1}=L^{\plpl}_{L,-1}=L^{\plpl}_{R,-1} = -\sqrt{2}m_\ell\sin\theta_\ell\, ,\\
& L^{\plmi}_{L,+1} = - L^{\mipl}_{R,-1} = -\sqrt{\frac{q^2}{2}}(1-\beta_\ell)(1-\cos\theta_\ell)\, ,\\
& L^{\mipl}_{L,+1} = -L^{\plmi}_{R,-1} = \sqrt{\frac{q^2}{2}}(1+\beta_\ell)(1+\cos\theta_\ell)\, ,\\
& L^{\plmi}_{R,+1}=-L^{\mipl}_{L,-1} = -\sqrt{\frac{q^2}{2}}(1+\beta_\ell)(1-\cos\theta_\ell)\, ,\\
& L^{\mipl}_{R,+1}=-L^{\plmi}_{L,-1} = \sqrt{\frac{q^2}{2}}(1-\beta_\ell)(1+\cos\theta_\ell)\, ,\\
& -L^{\plpl}_{L,0}=L^{\mimi}_{L,0}=-L^{\plpl}_{R,0}=L^{\mimi}_{R,0} = 2m_\ell\cos\theta_\ell\, ,\\
& L^{\plmi}_{L,0}=L^{\mipl}_{R,0} = \sqrt{q^2}(1-\beta_\ell)\sin\theta_\ell\, ,\quad L^{\mipl}_{L,0} = L^{\plmi}_{R,0} = \sqrt{q^2}(1+\beta_\ell)\sin\theta_\ell\, ,\\
& L^{\plpl}_{L,t} = L^{\mimi}_{L,t} = -L^{\plpl}_{R,t} = -L^{\mimi}_{R,t} = 2m_\ell\, .
 \end{align}
The combinations of $\lambda_1$ and $\lambda_2$ for which the amplitudes vanish are not shown.

 \subsection{$\Lambda\to N\pi$ decay amplitudes}
The $\Lambda_b \to \Lambda\ell^+\ell^-$ decay is followed by the subsequent parity violating weak decay of $\Lambda\to N\pi$ which is governed by the $\Delta S=1$ effective Hamiltonian
 \begin{equation}\label{eq:Heff2}
 \mathcal{H}^{\rm eff}_{\Delta S=1} = \frac{4G_F}{\sqrt{2}} V_{ud}^\ast V_{us} \big[ \bar{d}\gamma_\mu P_L u \big] \big[ \bar{u} \gamma^\mu P_L s\big]\, .
 \end{equation}  
The corresponding matrix elements are parametrized in terms of two hadronic parameters $\xi$ and $\omega$ as \cite{Boer:2014kda}
\begin{eqnarray}
	\mathcal{M}_2(s_k,s_N) &=& \langle p(k_1,s_N)\pi^-(k_2)\big| [ \bar{d}\gamma_\mu P_L u ][ \bar{u} \gamma^\mu P_L s ] \big| \Lambda(k,s_k)\rangle\, ,\nn\\
	&=& \bar{u}(k_1,s_N) ( \omega + \xi \gamma_5 ) u(k,s_k)\, .
\end{eqnarray}
The hadronic parameters $\xi,\omega$ can be extracted from the decay width and polarization measurements of $\Lambda\to p\pi^-$ decay. In terms of the kinematic variables the amplitudes can be written as \cite{Boer:2014kda}
\begin{eqnarray}
\begin{split}\label{eq:M2}
&\mathcal{M}_2\bigg(+\frac{1}{2},+\frac{1}{2}\bigg) = \bigg( \sqrt{r_+}\omega - \sqrt{r_-} \xi \bigg)\cos\frac{\theta_\Lambda}{2}\, ,\\
&\mathcal{M}_2\bigg(+\frac{1}{2},-\frac{1}{2}\bigg) = \bigg( \sqrt{r_+}\omega + \sqrt{r_-} \xi \bigg)\sin\frac{\theta_\Lambda}{2} e^{i\phi}\, ,\\
&\mathcal{M}_2\bigg(-\frac{1}{2},+\frac{1}{2}\bigg) = \bigg( -\sqrt{r_+}\omega + \sqrt{r_-} \xi \bigg)\sin\frac{\theta_\Lambda}{2} e^{-i\phi}\, ,\\
&\mathcal{M}_2\bigg(-\frac{1}{2},-\frac{1}{2}\bigg) = \bigg( \sqrt{r_+}\omega + \sqrt{r_-} \xi \bigg)\cos\frac{\theta_\Lambda}{2}\, ,
\end{split}
\end{eqnarray} 
where 
\begin{equation}
	r_\pm = (m_\Lambda \pm m_N)^2 - m_\pi^2\, .
\end{equation}
The total decay width of $\Lambda\to N\pi$ is given by
\begin{equation}
	\Gamma_\Lambda = \bigg(\frac{16G_F^2|V_{ud}^\ast V_{us}|^2}{2}\bigg) \frac{\sqrt{r_+r_-}}{16\pi m_\Lambda^3} \bigg( r_-|\xi|^2 + r_+|\omega|^2  \bigg)\, .
\end{equation}
For future convenience we introduce the parity violating parameter
\begin{equation}
	\alpha_\Lambda = \frac{-2\re(\xi\omega)}{\sqrt{\frac{r_-}{r_+}}|\xi|^2 + \sqrt{\frac{r_+}{r_-}}|\omega|^2 }\, .
\end{equation}

\section{Angular distribution \label{sec:angdist}} 
With the hadronic and leptonic amplitudes defined in the previous sections, we write down the four fold differential distribution of the four-body $\Lambda_b\to\Lambda(\to N\pi)\ell^+\ell^-$ decay
\begin{eqnarray}\label{eq:4fold}
\frac{d^4\mathcal{B}}{dq^2d\cos\theta_\ell d\cos\theta_\Lambda d\phi} &=& \frac{3}{8\pi} K(q^2,\cos\theta_\ell,\cos\theta_\Lambda,\phi)\, , 
\end{eqnarray} 
where $K(q^2,\cos\theta_\ell,\cos\theta_\Lambda,\phi)$ can be written in terms of a set of trigonometric functions and angular coefficients as
\begin{eqnarray}
K(q^2,\cos\theta_\ell,\cos\theta_\Lambda,\phi) &=& (K_{1ss} \sin^2\theta_\ell + K_{1cc} \cos^2\theta_\ell + K_{1c} \cos\theta_\ell )\, \nn\\
&+& (K_{2ss} \sin^2\theta_\ell + K_{2cc} \cos^2\theta_\ell + K_{2c} \cos\theta_\ell ) \cos\theta_\Lambda \, \nn\\
&+&( K_{3sc}\sin\theta_\ell\cos\theta_\ell + K_{3s}\sin\theta_\ell ) \sin\theta_\Lambda \sin\phi \, \nn\\
&+& ( K_{4sc}\sin\theta_\ell\cos\theta_\ell + K_{3s}\sin\theta_\ell ) \sin\theta_\Lambda \cos\phi \, .
\end{eqnarray}
As the final state leptons are massive, each of the angular coefficients can be decomposed as
\begin{equation}
 K_{\{\cdots\}} = \mathcal{K}_{\{\cdots\}} + \frac{m_\ell}{\sqrt{q^2}} \mathcal{K}_{\{\cdots\}}^\prime + \frac{m_\ell^2}{q^2}\mathcal{K}_{\{\cdots\}}^{\prime\prime}\, ,
\end{equation}
where $\{\cdots\}$ correspond to the suffixes $1ss, 1cc, 1c, 2ss,2cc,2c,3sc,3s,4sc,4s$. The first term $\mathcal{K}$ correspond to the results when the leptons are massless while the linear and the quadratic mass corrections are given by $\mathcal{K}^{\prime}$ and $\mathcal{K}^{\prime\prime}$ respectively. Deferring the derivations and the detailed expression of $\mathcal{K}$, $\mathcal{K}^{\prime}$ and $\mathcal{K}^{\prime\prime}$ to Appendix \ref{sec:angular}, we note the following points.
\begin{enumerate}
\item[\textbullet] The interference of the SM amplitudes with the scalar amplitudes ($A_{\rm S\perp,\|}, A_{\rm P\perp,\|}$) appear in $\mathcal{K}^{\prime,\prime\prime}$ only. Therefore in the massless limit the interference terms vanish.
\item[\textbullet] There are no linear mass corrections to $K_{3sc}$ and $K_{4sc}$ \emph{i.e.,} $\mathcal{K}^\prime_{3sc}=0, \mathcal{K}^\prime_{4sc}=0$.
\item[\textbullet] There are  no quadratic mass corrections to $K_{1c}, K_{2c}, K_{3s}$ and $K_{4s}$ \emph{i.e.,} $\mathcal{K}^{\prime\prime}_{1c,2c,3s,4s}=0$. The linear mass corrections also vanish in the SM \emph{i.e.,} $\mathcal{K}^{\prime}_{1c,2c,3s,4s}=0$.
\item[\textbullet] There is no SP contribution to $K_{3sc}$ and $K_{4sc}$. These angular coefficients are therefore not sensitive to $\mC_{S,P}^{(\prime)}$ couplings.
\item[\textbullet] There is no pseudo-scalar contribution ($A_{\rm P\|,P\perp}$) to $K_{1c}$, $K_{2c}$, $K_{3s}$ and $K_{4s}$. Therefore, these angular coefficients are not sensitive to $\mC_{P}^{(\prime)}$.
\end{enumerate}

All the long- and short-distance physics in the SM and beyond are essentially contained in the angular coefficients. These coefficients therefore need to be measured in experiments.  Usually, one constructs observables through weighted average of the differential distributions Eq.~(\ref{eq:4fold}) over the angular variables $\theta_\ell, \theta_\Lambda$ and $\phi$ which can be measured in experiments
\begin{equation}
	X(q^2) = \displaystyle\int \frac{d^4\mathcal{B}}{dq^2 d\cos\theta_\ell d\cos\theta_\Lambda d\phi} \omega_X (q^2, \cos\theta_\ell \cos\theta_\Lambda, \phi) d\cos\theta_\ell d\cos\theta_\Lambda d\phi\, .
\end{equation}
The observables that we will consider in this paper are listed below.
\begin{enumerate}
\item[\textbullet] The differential branching ratio is obtained with the weight $\omega_X = 1$
\begin{equation}
	\frac{d\mathcal{B}}{dq^2} = 2K_{1ss} + K_{1cc}\, .
\end{equation} 

\item[\textbullet] The fraction of longitudinal fraction is 
\begin{equation}
F_L = \frac{ 2K_{1ss} - K_{1cc} }{ 2K_{1ss} + K_{1cc} }\, ,\quad \omega_{F_L} = \frac{ 2 - 5\cos^2\theta_\ell }{d\mathcal{B}/dq^2}\, .
\end{equation} 

\item[\textbullet] In addition to defining the well known lepton side forward-backward asymmetry\footnote{With this definition the sign of $A^\ell_{\rm FB}$ is opposite to what we had in the earlier paper \cite{Das:2018sms}.}
\begin{equation}
	A^\ell_{\rm FB} = \frac{3}{2} \frac{K_{1c}}{ 2K_{1ss} + K_{1cc} }\, ,\quad\quad \omega_{A^\ell_{\rm FB}} = \frac{{\rm sgn}[\cos\theta_\ell]}{d\mathcal{B}/dq^2}\, ,
\end{equation}
we discuss the hadronic side forward-backward asymmetry $A^\Lambda_{\rm FB}$ and a combined forward-backward asymmetry $A^{\ell\Lambda}_{\rm FB}$ which are defined as
\begin{eqnarray}
 A^\Lambda_{\rm FB} &=& \frac{1}{2} \frac{ 2K_{2ss} + K_{2cc} }{ 2K_{1ss} + K_{1cc} }\, ,\quad\quad \omega_{A^\Lambda_{\rm FB}} = \frac{{\rm sgn}[\cos\theta_\Lambda]}{d\mathcal{B}/dq^2}\, ,\\
 A^{\ell\Lambda}_{\rm FB} &=& \frac{3}{4} \frac{ K_{2c} }{ 2K_{1ss} + K_{1cc} }\, ,\quad\quad \omega_{A^{\ell\Lambda}_{\rm FB}} = \frac{{\rm sgn}[\cos\theta_\ell\cos\theta_\Lambda]}{d\mathcal{B}/dq^2}\, .
\end{eqnarray}

\item[\textbullet] We also analyze the angular observables
\begin{eqnarray}
\hat{K}_{4sc} &=& \frac{K_{4sc}}{2K_{1ss}+K_{1cc}}\, ,\\
\hat{K}_{4s} &=& \frac{K_{4s}}{2K_{1ss}+K_{1cc}}\, .
\end{eqnarray}

\item[\textbullet] We also define the following ratio that is sensitive to LFU violation
\begin{equation}
R^{\tau/e}_{\Lambda_b} = \frac{\displaystyle\int_{q^2_{\rm min}}^{q^2_{\rm max}} dq^2 d\mathcal{B}(\Lambda_b\to\Lambda(\to p\pi)\tau^+\tau^-)/dq^2 }{\displaystyle\int_{q^2_{\rm min}}^{q^2_{\rm max}} dq^2 d\mathcal{B}(\Lambda_b\to\Lambda(\to p\pi) e^+e^-)/dq^2 }\, .
\end{equation} 
\end{enumerate}
When we study $R^{\tau/e}_{\Lambda}$ in NP, we assume that the electron mode is SM like. 

In the next sections we study these observables as functions of the dielpton invariant mass squared for $\Lambda_b\to \Lambda(\to p\pi)\tau^+\tau^-$ decay. We also make binned average predictions in different $q^2$ bins.

\section{$\Lambda_b \to \Lambda$ form factors \label{sec:ff}}
The $\Lambda_b \to \Lambda$ hadronic matrix elements corresponding to the operators (\ref{eq:Heff1}) are parametrized in terms of ten $q^2$ dependent helicity form factors $f^V_{t,0,\perp}$, $f^A_{t,0,\perp}$, $f^T_{0,\perp}$, $f^{T5}_{0,\perp}$ in Appendix \ref{sec:hme}. For our numerical analysis we take the form factors from the calculation in lattice QCD \cite{Detmold:2016pkz} \footnote{The label of the form factors in \cite{Detmold:2016pkz} is different from ours. The relations are $f^V_{t,0,\perp} = f_{0,+,\perp}$, $f^A_{t,0,\perp} = g_{0,+,\perp}$, $f^T_{0,\perp}=h_{+,\perp}$ and $f^{T5}_{0,\perp}=\tilde{h}_{+,\perp}$. }. To obtain the $q^2$ dependence and estimate the uncertainties, the lattice calculations are fitted to two $z$-parameterizations. In the so called ``nominal" fit the parametrization read
\begin{equation}\label{eq:znominal}
f(q^2) = \frac{1}{1-q^2/(m^{f}_{\rm pole})^2} \big[ a^{f}_0 + a_1^{f} z(q^2,t_+) \big]\, .
\end{equation}
The statistical uncertainties of the observable are calculated from the ``nominal'' fit, while, to estimate the systematic uncertainties a ``higher-order'' fit is used for which the fit function is
\begin{equation}\label{eq:zhigher}
f(q^2) = \frac{1}{1-q^2/(m^{f}_{\rm pole})^2} \big[ a^{f}_0 + a_1^{f} z(q^2,t_+) + a_2^{f} (z(q^2,t_+))^2 \big]\, ,
\end{equation}
where $z(q^2,t_+)$ is defined as
\begin{equation}
	z(q^2,t_+) = \frac{ \sqrt{t_+-q^2} - \sqrt{t_+-t_0}  }{ \sqrt{t_+-q^2} + \sqrt{t_+-t_0} }\, ,
\end{equation}
with $t_0 = (\mLb-\mL)^2$, $t_+=(m_B+m_K)^2$. The values of the fit parameters and all the masses are taken from \cite{Detmold:2016pkz}.

The method to estimate the central values and the statistical and systematic uncertainties for any observable $\mathcal{O}$ is as follows.
\begin{enumerate}
	\item[\textbullet] The central value and the uncertainty obtained using the nominal fit Eq.~(\ref{eq:znominal}) is denoted as
	\begin{equation}
		O\, ,\quad \sigma_{O}\, .
	\end{equation}
	\item[\textbullet] The central value and the uncertainty computed using the higher order fit Eq.~(\ref{eq:zhigher}) is denoted as
	 \begin{equation}
	 O_{\rm HO}\, ,\quad \sigma_{O,{\rm HO}}\, .
	 \end{equation}
	 \item[\textbullet] The central value, the statistical uncertainty, and the systematic uncertainty are then given by
	 \begin{equation}
	 O \pm \sigma_{O,\rm stat} \pm \sigma_{O,\rm syst}\, ,
	 \end{equation}
	 where the statistical uncertainty and the systematic uncertainty are given by
	 \begin{eqnarray}
	 	\sigma_{O,\rm stat} &=& \sigma_{O}\, ,\nn\\
	 	\sigma_{O,\rm syst} &=& {\rm max}\bigg( |O_{\rm HO} - O|, \sqrt{|\sigma^2_{O,\rm HO} - \sigma^2_{O}|}  \bigg)\, .
	 \end{eqnarray}
\end{enumerate}

Unless explicitly mentioned, we limit the range of the $q^2$ to $14.18 < q^2 < (\mLb - \mL)^2$ GeV$^2$. For simplicity of the analysis we neglect the correlations among the fit parameters given in \cite{Detmold:2016pkz}.

\section{Numerical Analysis \label{sec:numerical}}
In this section we study the $\Lambda_b\to(\Lambda\to p\pi)\tau^+\tau^-$ observables in the SM and in NP. We note that so far no $b\to s\tau^+\tau^-$ transition has yet been observed. Only the LHCb has searched for $B_s\to\tau^+\tau^-$ and provides an upper limit \cite{Aaij:2017xqt}
\begin{equation}
\mathcal{B}(B_s\to\tau^+\tau^-)^{\rm LHCb} < 6.8 \times 10^{-3}\, ,
\end{equation}
whereas the upper limit on $B^+\to K^+\tau^+\tau^-$ from BaBar \cite{TheBaBar:2016xwe} is
\begin{equation}
\mathcal{B}(B^+\to K^+\tau^+\tau^-)^{\rm BaBar} < 2.25 \times 10^{-3}\, .
\end{equation} 
The LHCb is capable to search for $B\to K^\ast\tau^+\tau^-$ and $B_s\to\phi\tau^+\tau^-$. On the other hand, the upcoming Belle II which is expected to run at $\Upsilon(4s)$ will be able to study $B\to K^{(\ast)}\tau^+\tau^-$. Due partly to the unsatisfactory experimental situation, the $b\to s\tau^+\tau^-$ has received very little theoretical attention. Recently, interest in $b\to s\tau^+\tau^-$ has been revived due to the observations in  Refs.~\cite{Alonso:2015sja,Crivellin:2017zlb,Capdevila:2017iqn} that an explanation of the anomalies in the $b\to c\ell\nu$ and $b\to s\mu^+\mu^-$ modes lead to large enhancements to $b\to s\tau^+\tau^-$ rates. 

\begin{figure}[h!]
	\begin{center}
		\includegraphics[scale=0.3]{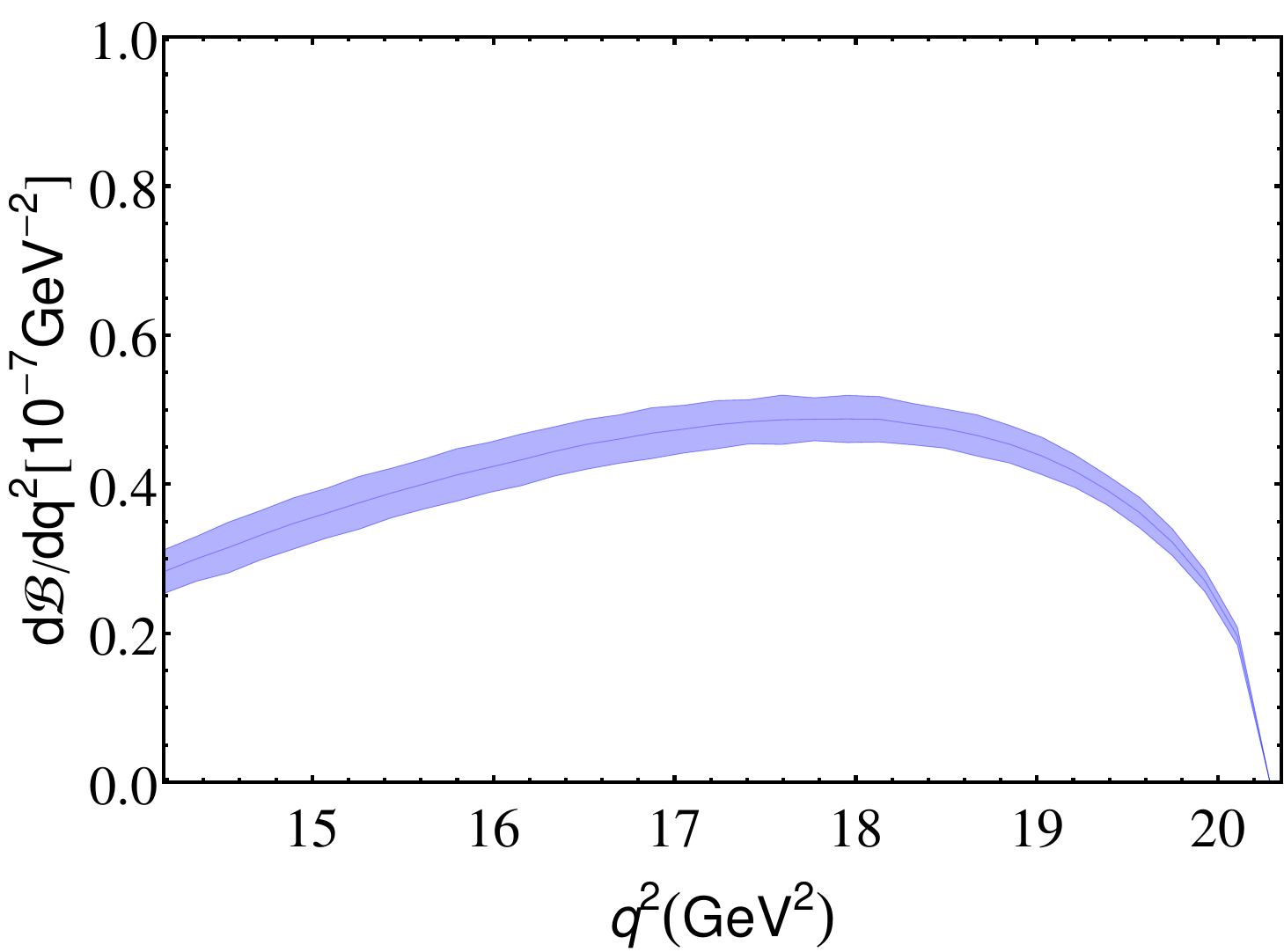}
		\includegraphics[scale=0.3]{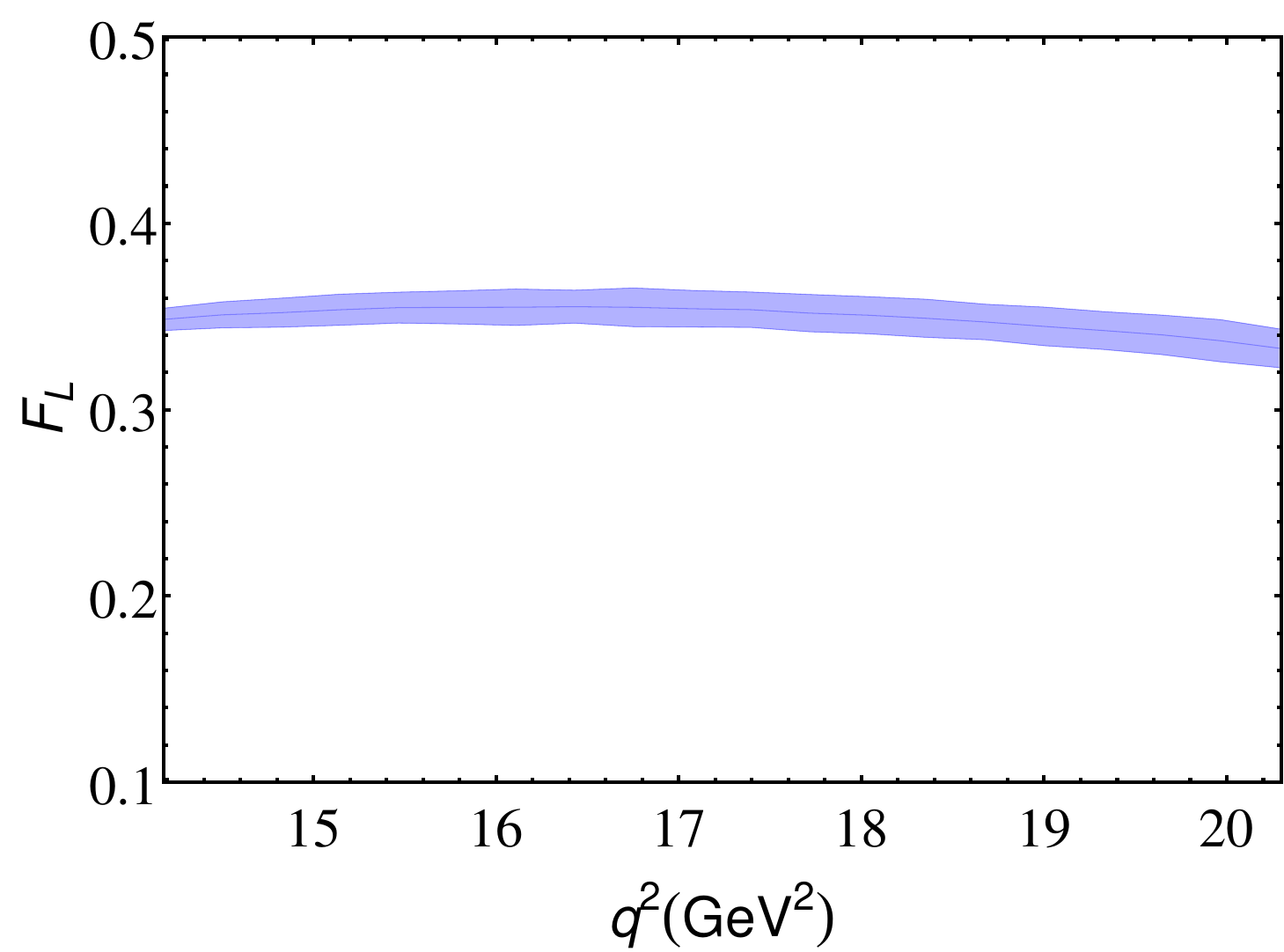}
		\includegraphics[scale=0.3]{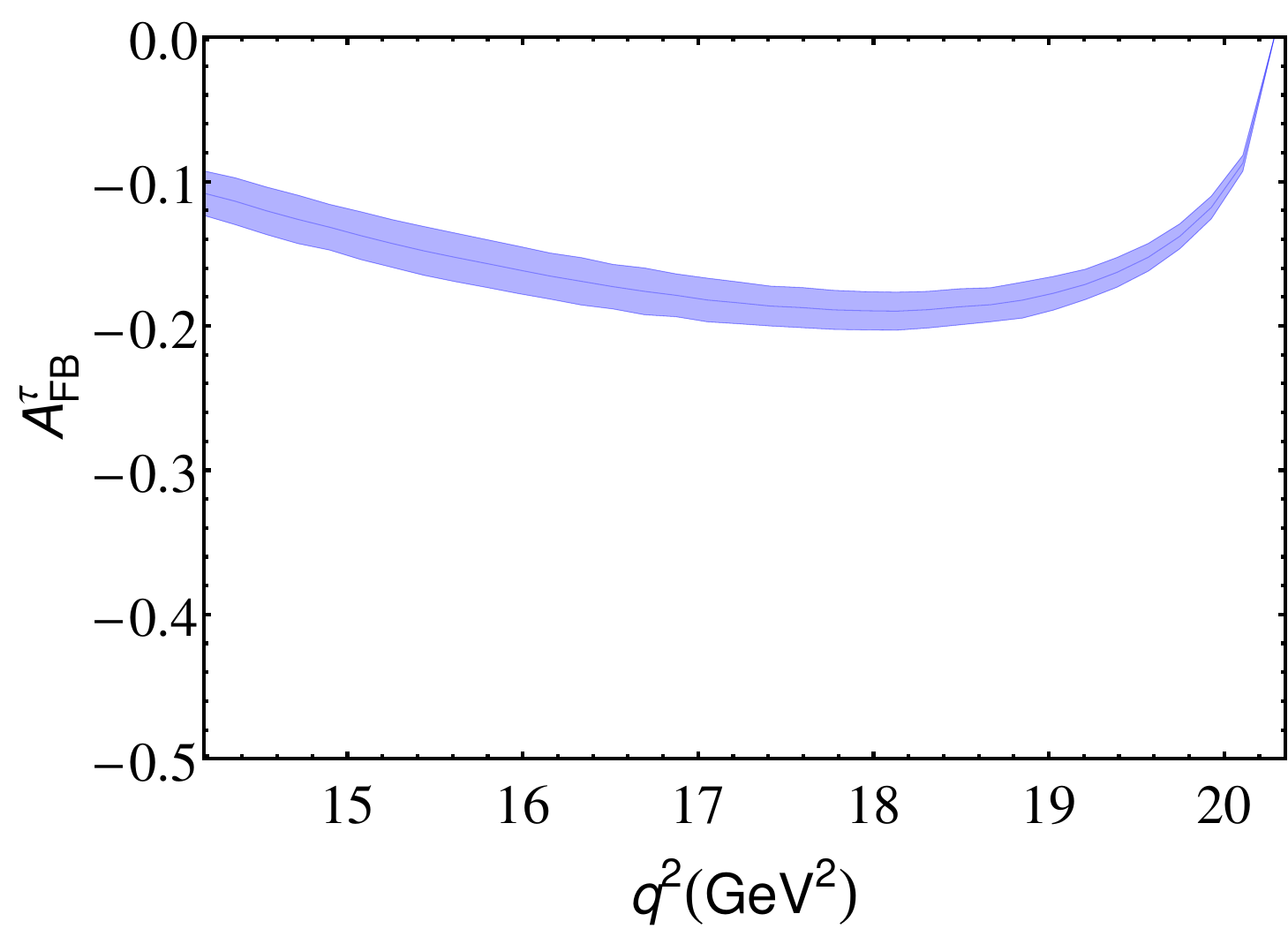}
		\includegraphics[scale=0.3]{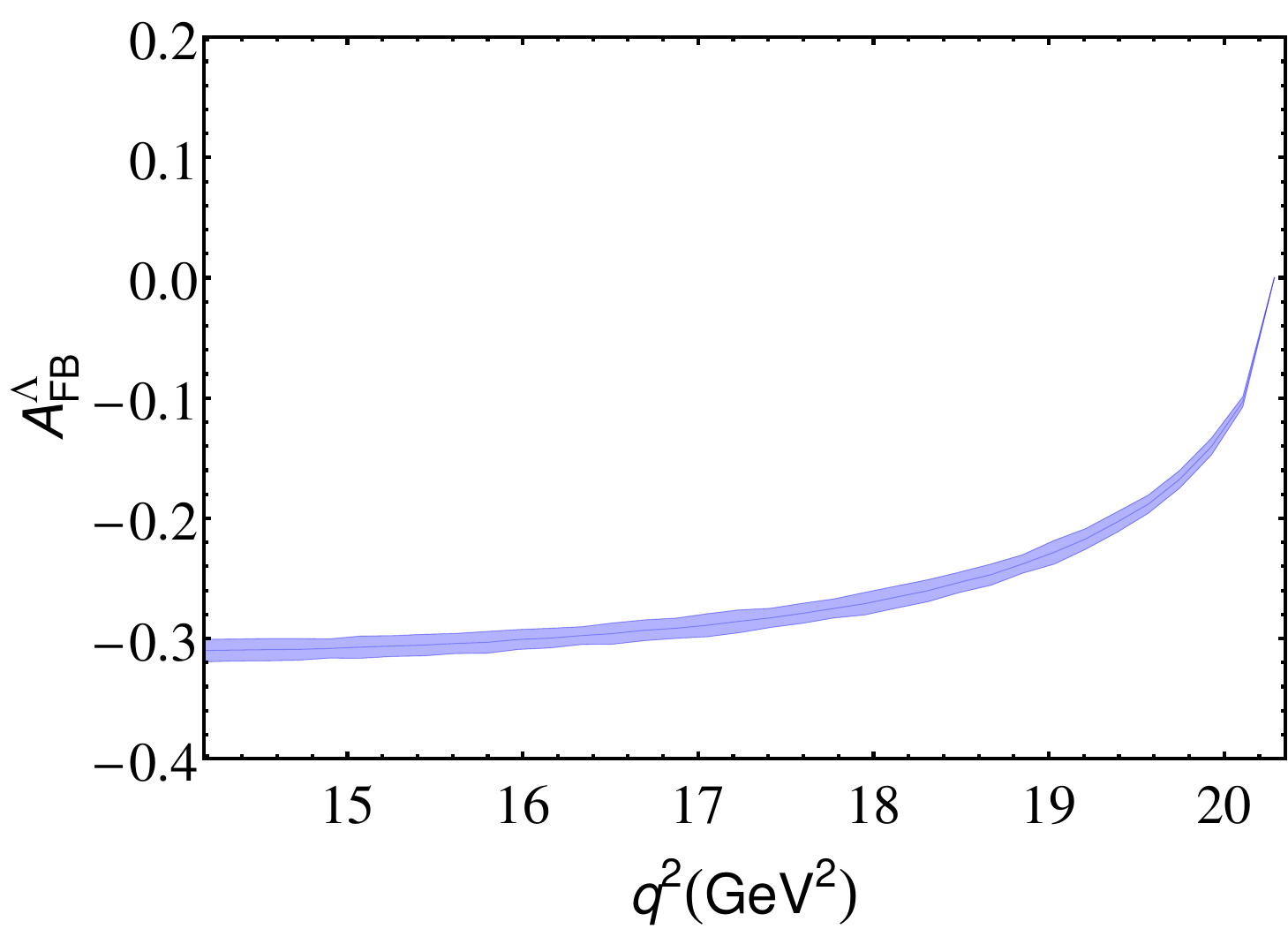}
		\includegraphics[scale=0.3]{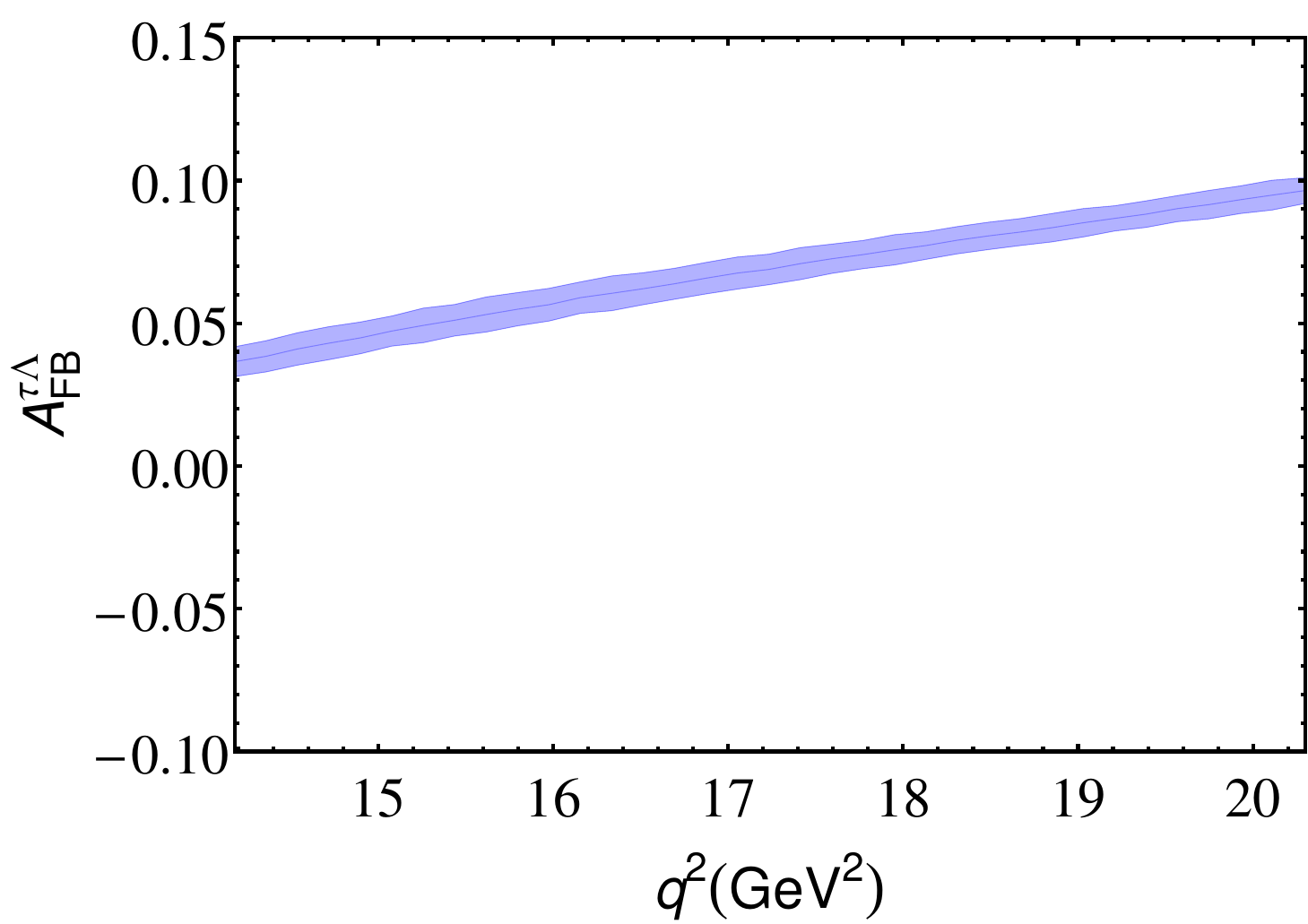}
		\includegraphics[scale=0.3]{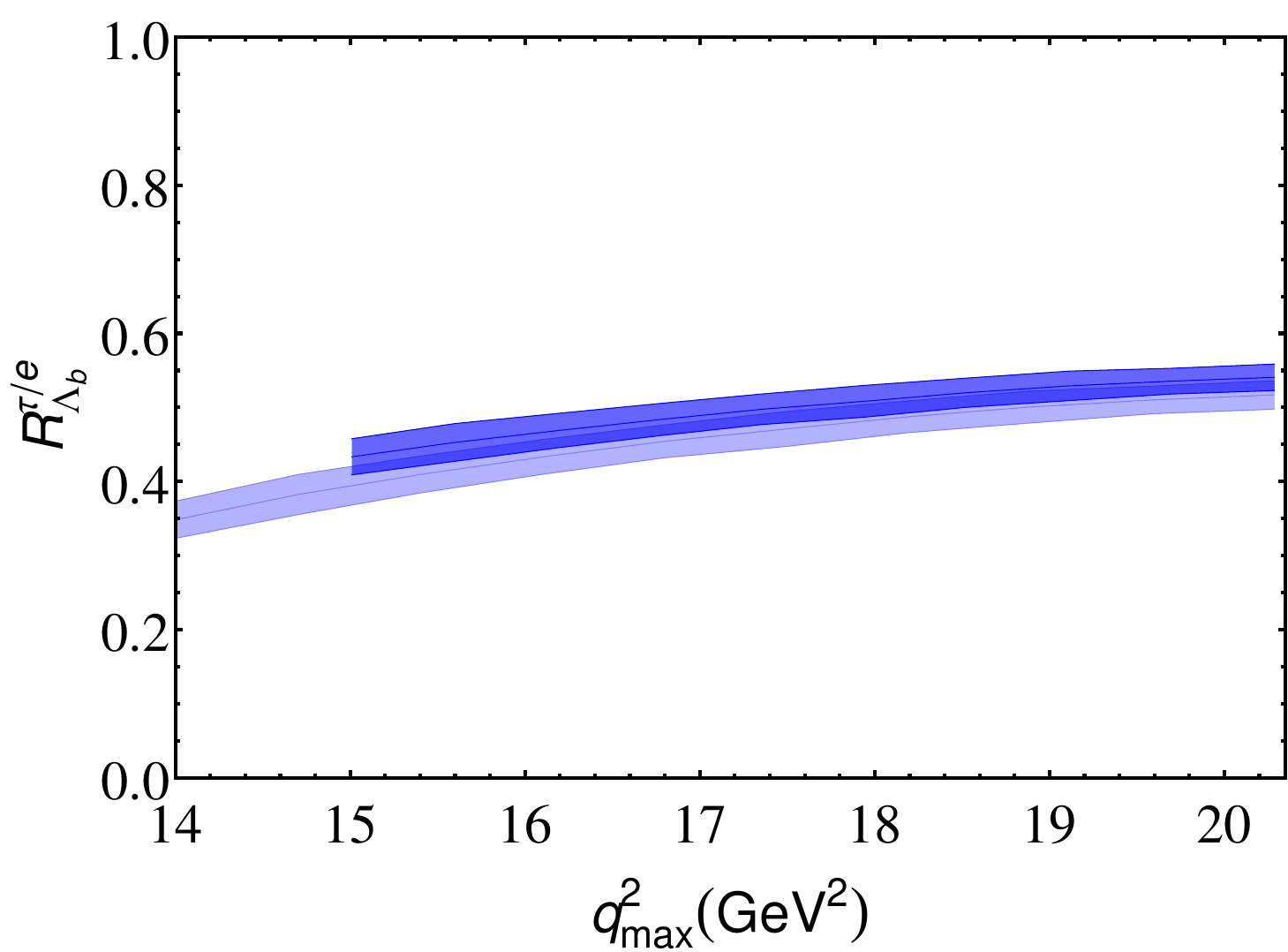}
		\caption{The SM predictions of various $\Lambda_b\to\Lambda(\to p^+\pi^-)\tau^+\tau^-$ observables as a function of $q^2$. The $R^{\tau/e}_\Lambda$ is shown as a function of the upper limit of integration $q^2_{\rm max}$ for two values of lower integration limit $q^2_{\rm min}=14$ GeV$^2$ (light blue) and $q^2_{\rm min}=15$ GeV$^2$ (dark blue). The bands correspond to the uncertainties discussed in text. Since the OPE at large $q^2$ does not capture local resonance structures, the distributions can be locally off from the OPE predictions.\label{fig:SM}}
	\end{center}
\end{figure}
\begin{figure}[h!]
	\begin{center}
		\includegraphics[scale=0.3]{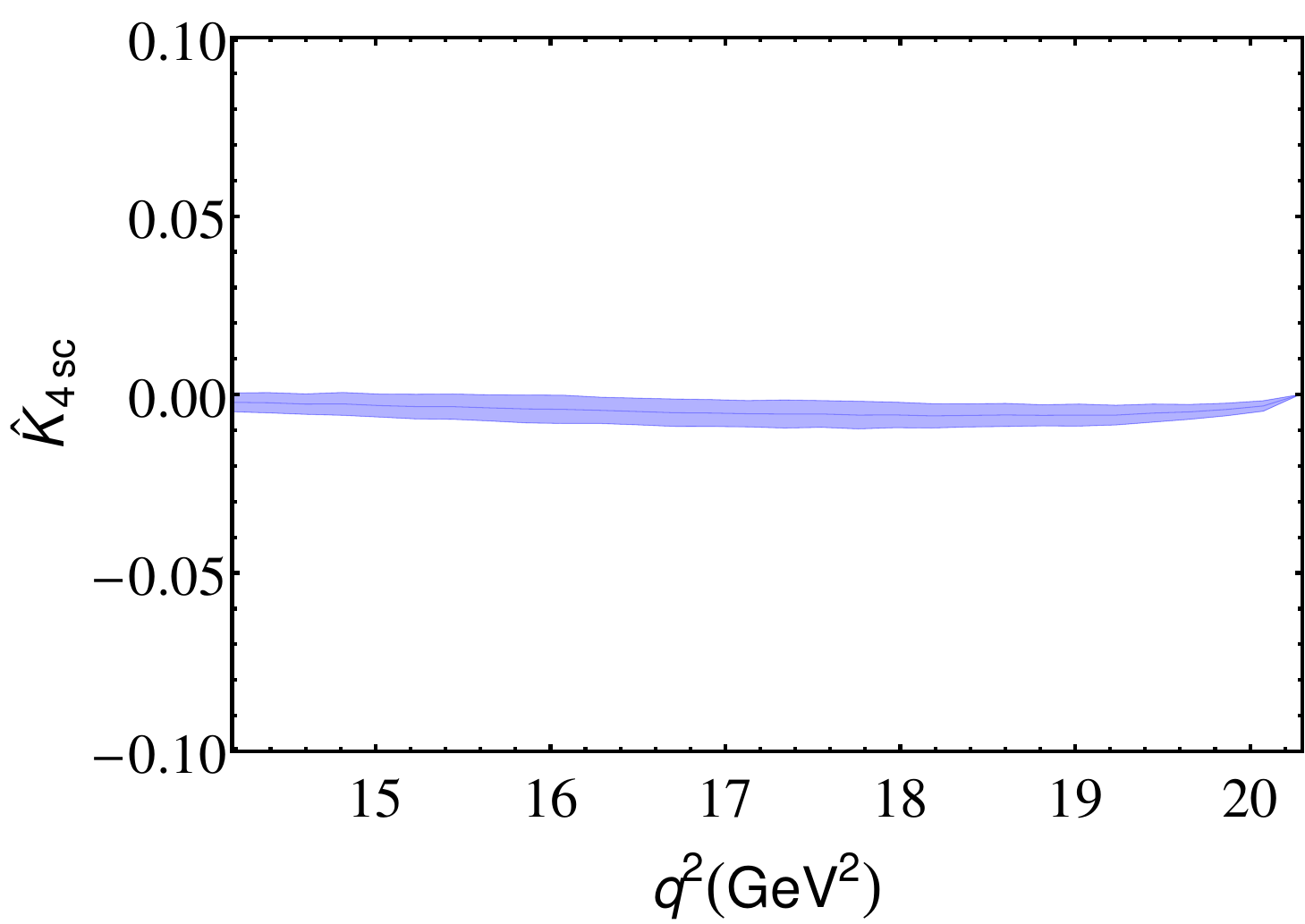}
		\includegraphics[scale=0.3]{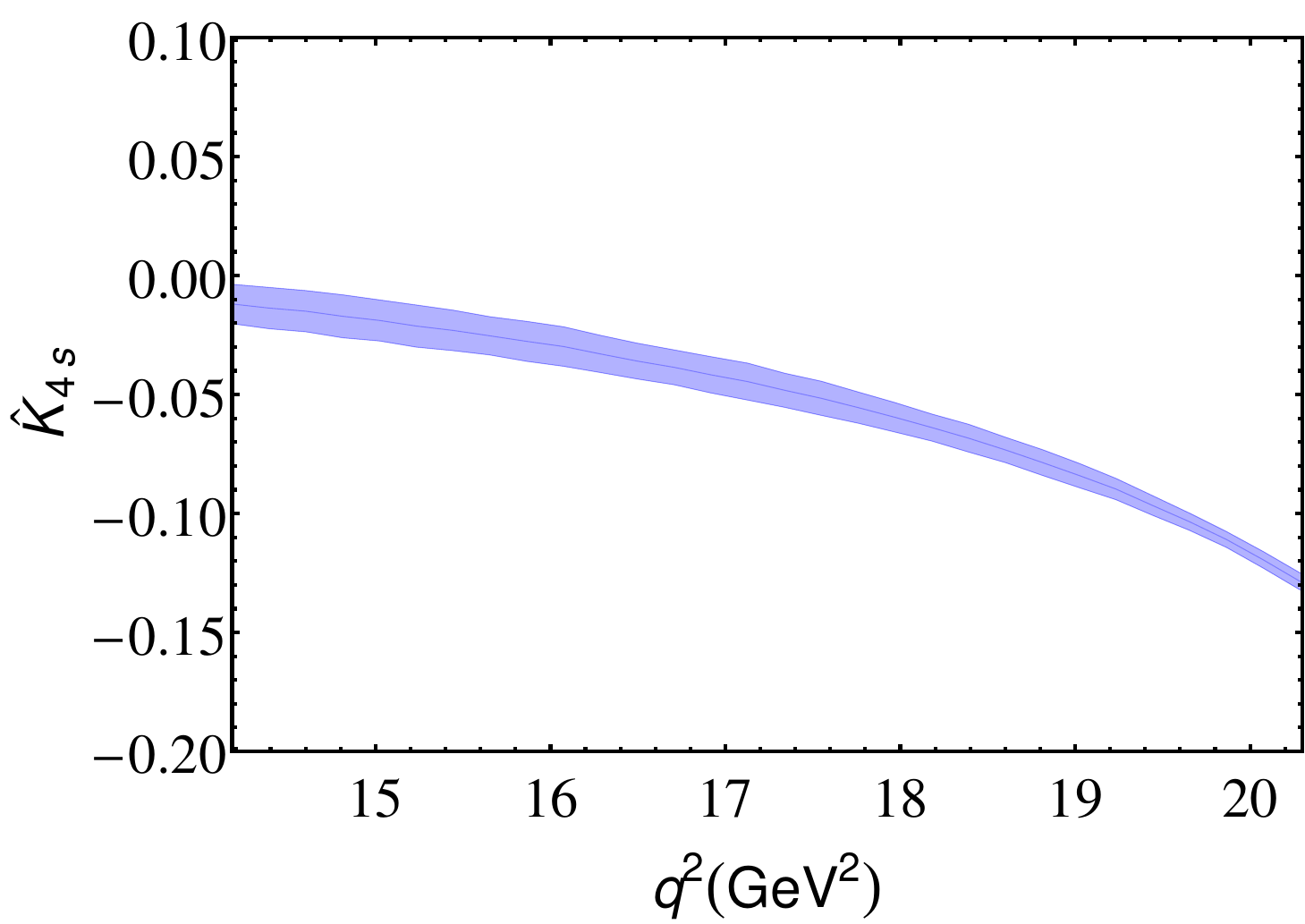}
		\caption{The SM predictions of $\hat{K}_{4sc}$ and $\hat{K}_{4s}$ for $\Lambda_b\to\Lambda(\to p^+\pi^-)\tau^+\tau^-$ as a function of $q^2$. The bands correspond to the uncertainties discussed in text. Since the OPE at large $q^2$ does not capture local resonance structures, the distributions can be locally off from the OPE predictions. \label{fig:SM2}}
	\end{center}
\end{figure}
For the SM analysis we set the couplings $\mC_{V,A}^{(\prime)} = 0, \mC_{S,P}^{(\prime)} = 0$. The SM Wilson coefficient $\mC_{7,9}^{\rm eff}$ (see Appendix \ref{app:C910}) whose dominant contributions come from the short-distance corrections to $\mathcal{O}_{7,9}$, also contain the leading contributions from the factorizable non-local matrix elements of the purely hadronic operators $\mathcal{O}_{1-6,8}$ with quark electro-magnetic current \cite{Grinstein:2004vb,Beylich:2011aq}. The sub-leading contributions from an operator product expansion (OPE) \cite{Beylich:2011aq} arise from dimension-5 operators, whose matrix elements are suppressed at high-$q^2$ by $\Lambda_{\rm had}^2/Q^2 \sim 2\%$, where $Q^2\sim\{q^2, m_b^2 \}$. We ignore the yet to be derived non-factorizable spectator corrections for baryonic decay which are expected to be significant only at the low-$q^2$ region. However, the perturbative expressions of $\mC_{7,9}^{\rm eff}$ given in Appendix \ref{app:C910} do not accurately describe the effects of multiple broad charmonium resonances at high-$q^2$ \cite{Lyon:2014hpa}, which lead to the violation of quark-hadron duality. In the OPE based approach, these are beyond the neglected orders in $\alpha_s$ and $\Lambda_{\rm had}/Q$. The duality violation in $B\to K\ell^+\ell^-$ was estimated to be around 2\% for the decay rate integrated over the high-$q^2$ region \cite{Beylich:2011aq}. Since a detailed discussion of quark-hadron duality violation in $\Lambda_b\to \Lambda(\to p\pi)\tau^+\tau^-$ is beyond the scope of this paper, following \cite{Detmold:2016pkz} we include 5\% corrections to both $\mC_{7,9}^{\rm eff}$ in our analysis to account for this effect. In addition to this correction, the $q^2$ distributions of observables, that we discuss next, come with a caveat; since the non-perturbative effects of local resonance structures are not captured by the OPE at any order in perturbation expansion, the distributions can be locally away from the OPE predictions by large amount. Even for observables integrated over the entire low recoil region, the effects can still be significant. 

The central values and uncertainties of some of the inputs including the value of the parity violating parameter $\alpha_\Lambda$ are collected in table \ref{tab:inputs} in Appendix \ref{sec:inputs} while the rest are discussed in the text. The $q^2$ distributions of the SM differential branching ratio, the longitudinal polarization fraction $F_L$, and three angular asymmetries $A_{\rm FB}^{\tau,\Lambda,\tau\Lambda}$ are shown in figure \ref{fig:SM}. To obtain the plots the lattice QCD form factors are employed as discussed in Sec.~\ref{sec:ff}. The bands correspond to the uncertainties of the form factors and other inputs. From the plot of the differential branching ratio $d\mathcal{B}/dq^2$ we see that the branching ratio is reduced compared to $\Lambda_b\to\Lambda\mu^+\mu^-$ by phase space suppression factor $\beta_\ell$. For the angular asymmetries $A^{\tau}_{\rm FB}$ ($A^{\tau\Lambda}_{\rm FB}$) the mass corrections terms to $K_{1c}$ ($K_{2c}$), \emph{i.e.,} $\mathcal{K}_{1c}^{\prime,\prime\prime}$ ($\mathcal{K}_{2c}^{\prime,\prime\prime}$) vanish in the SM. Therefore the effect of lepton mass mainly come due to normalization by the differential decay width and the factor $\beta_\ell$ appearing in $\mathcal{K}_{1c}$ ($\mathcal{K}_{2c}$). 
In the last panel of Fig.~\ref{fig:SM} we have shown $R^{\tau/e}_{\Lambda_b}$ as a function of $q^2_{\rm max}$ for two values of $q^2_{\rm min}$. In figure \ref{fig:SM2} we show the angular coefficients $\hat{K}_{4sc}$ and $\hat{K}_{4s}$. For the observable $\hat{K}_{4s}$, in the numerator the mass correction term $\mathcal{K}_{4s}^{\prime\prime}=0$ and $\mathcal{K}_{4s}^{\prime}=0$ in the SM. The effect of lepton mass therefore comes due to the normalization by the branching ratio and through the factor $\beta_\ell$ that appear in $\mathcal{K}_{4s}$.

	\begin{table*}\centering
		\begin{tabular}{l c c c c  }\hline \hline
			$[q^2_{\rm min},q^2_{\rm max}]$	& $\langle \mathcal{B}\rangle \times 10^{7}$ & $\langle F_L\rangle$ & $\langle A_{\rm FB}^\tau\rangle$ & $\langle A_{\rm FB}^\Lambda\rangle$  \\\hline
			[14,15]	 & $ 0.31 \pm 0.03 $ & $ 0.350\pm 0.007 $ & $-0.120\pm 0.016$ & $-0.309\pm 0.009$  \\\ 
			[15,16]	 & $ 0.39 \pm 0.03 $ & $ 0.354\pm 0.009 $ & $-0.149\pm 0.017$ & $-0.304\pm 0.009$  \\\
			[16,18]	 & $ 0.93 \pm 0.06 $ & $ 0.353\pm 0.010 $ & $-0.180\pm 0.015$ & $-0.289\pm 0.009$  \\\
			[18,20]	 & $ 0.83 \pm 0.05 $ & $ 0.346\pm 0.010 $ & $-0.173\pm 0.010$ & $-0.227\pm 0.009$ \\\
			[15,20]	 & $ 2.16 \pm 0.13 $ & $ 0.351\pm 0.009 $ & $-0.171\pm 0.014$ & $-0.268\pm 0.008$ \\
			\hline\hline
		\end{tabular}\caption{The total integrated branching ratio and the values of the observables in different $q^2$ bins (in GeV$^2$). The errors correspond to the uncertainties discussed in text. \label{table:ObsSMtau}}
\end{table*}
%
\begin{table*}\centering
	\begin{tabular}{l c c c }\hline \hline
		$[q^2_{\rm min},q^2_{\rm max}]$	& $\langle A^{\tau\Lambda}_{\rm FB}\rangle$ & $\langle \hat{K}_{4sc}\rangle$ & $\langle \hat{K}_{4s}\rangle$   \\\hline
		[14,15]	 & $ 0.041 \pm 0.005 $ & $ -0.0026\pm 0.0030 $ & $-0.0148\pm 0.0086$ \\\ 
		[15,16]	 & $ 0.052 \pm 0.006 $ & $ -0.0037\pm 0.0036 $ & $-0.0243\pm 0.0089$  \\\
		[16,18]	 & $ 0.067 \pm 0.006 $ & $ -0.0053\pm 0.0035 $ & $-0.0446\pm 0.0075$  \\\
		[18,20]	 & $ 0.084 \pm 0.005 $ & $ -0.0053\pm 0.0028 $ & $-0.0827\pm 0.0050$  \\\
		[15,20]	 & $ 0.071 \pm 0.005 $ & $ -0.0050\pm 0.0031 $ & $-0.0055\pm 0.0067$  \\
		\hline\hline
	\end{tabular}\caption{The values of the observables in different $q^2$ bins (in GeV$^2$). The errors correspond to the uncertainties discussed in text. \label{table:ObsSMtau2}}
\end{table*}

In tables \ref{table:ObsSMtau} and \ref{table:ObsSMtau2} we present the values of the observables in different di-lepton invariant mass squared bins. To be more precise, we have shown the total integrated branching ratio which defined as
\begin{equation}
	\langle \mathcal{B} \rangle = \displaystyle\int_{q^2_{\rm min}}^{q^2_{\rm max}} dq^2 \frac{d \mathcal{B}}{dq^2}\, .
\end{equation}
The binned averaged values of an observables such as $\langle F_L \rangle$ is defined as 
\begin{equation}
	\langle F_L \rangle = \frac{\displaystyle\int_{q^2_{\rm min}}^{q^2_{\rm max}} dq^2 (2K_{1ss} - K_{1cc}) }{\displaystyle\int_{q^2_{\rm min}}^{q^2_{\rm max}} dq^2 (2K_{2ss} + K_{2cc}) }\, .
\end{equation}
The $q^2$-integrated values of $R^{\tau/e}_{\Lambda_b}$ for two different low recoil bins are given below
\begin{eqnarray}
	\begin{split}
R^{\tau/e}_{\Lambda_b} &=& 0.5077 \pm 0.0210\,  ,\quad [14-(\mLb-\mL)^2]~ \text{GeV}^2\, ,\\
R^{\tau/e}_{\Lambda_b} &=& 0.5315 \pm 0.0189\,  ,\quad [15-(\mLb-\mL)^2]~ \text{GeV}^2\, .
    \end{split}
\end{eqnarray}

\begin{figure}[h!]
	\begin{center}
		\includegraphics[scale=0.4]{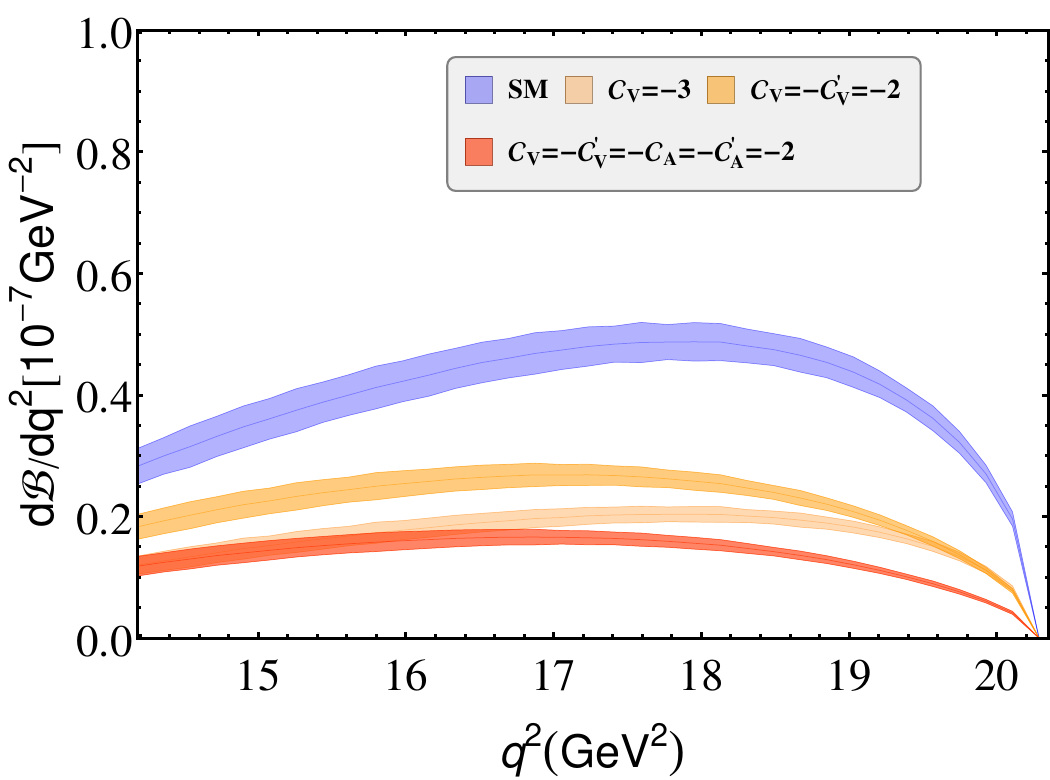}
		\includegraphics[scale=0.4]{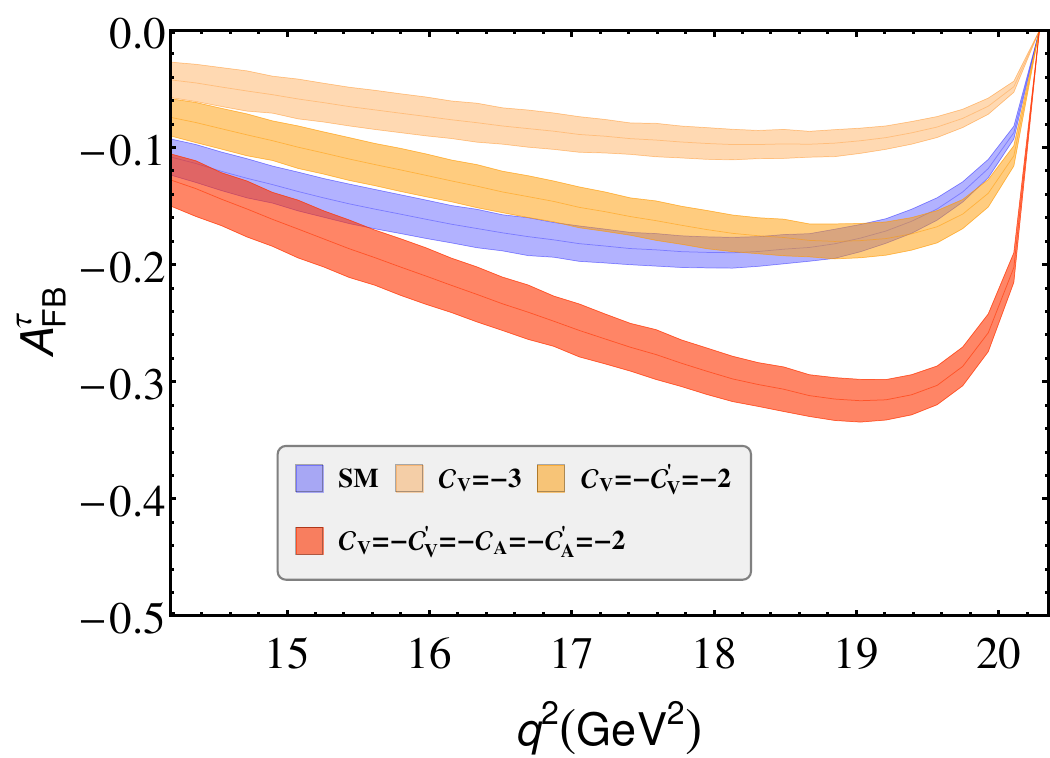}
		\includegraphics[scale=0.4]{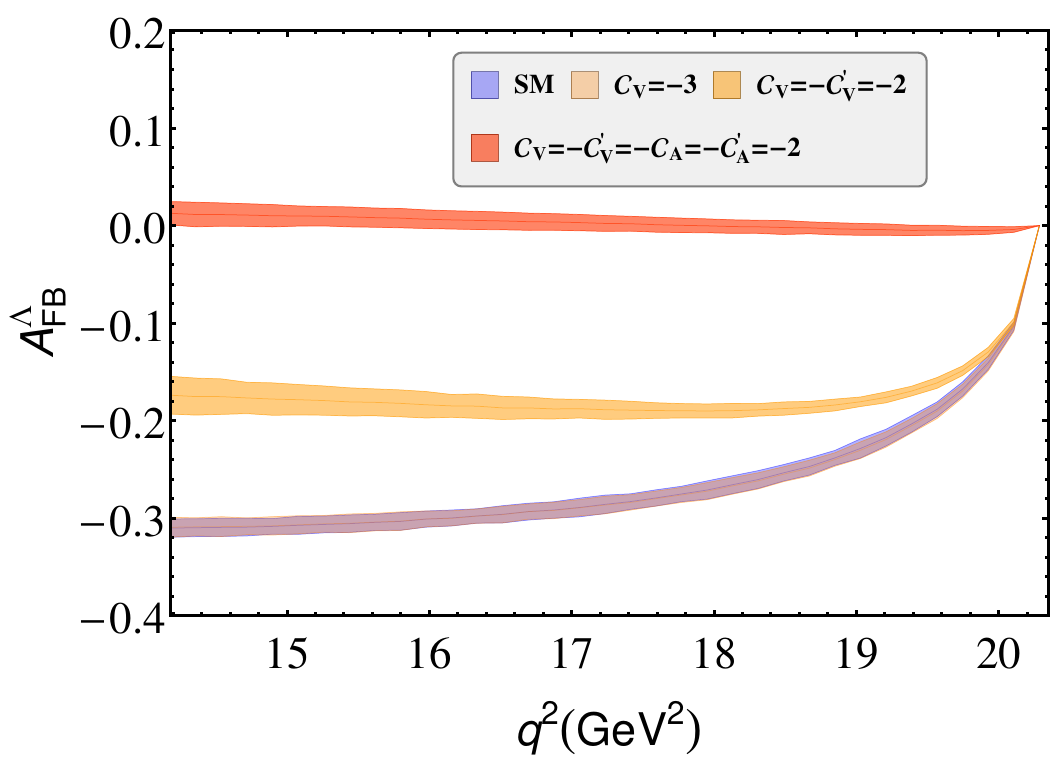}
		\includegraphics[scale=0.4]{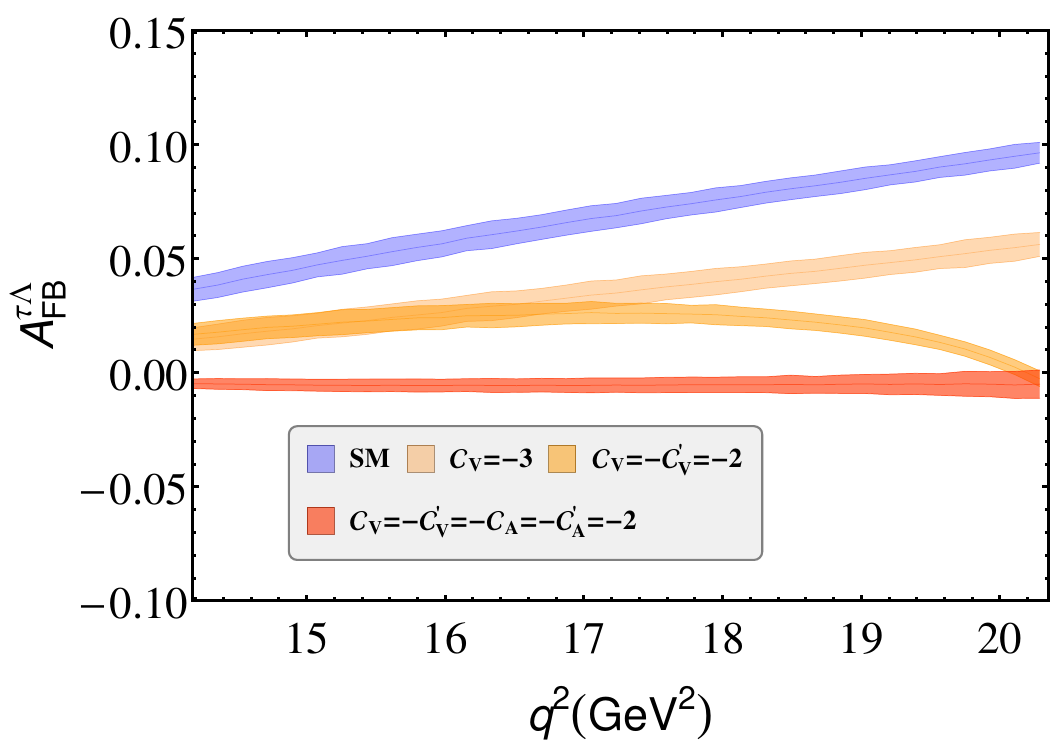}
		\includegraphics[scale=0.41]{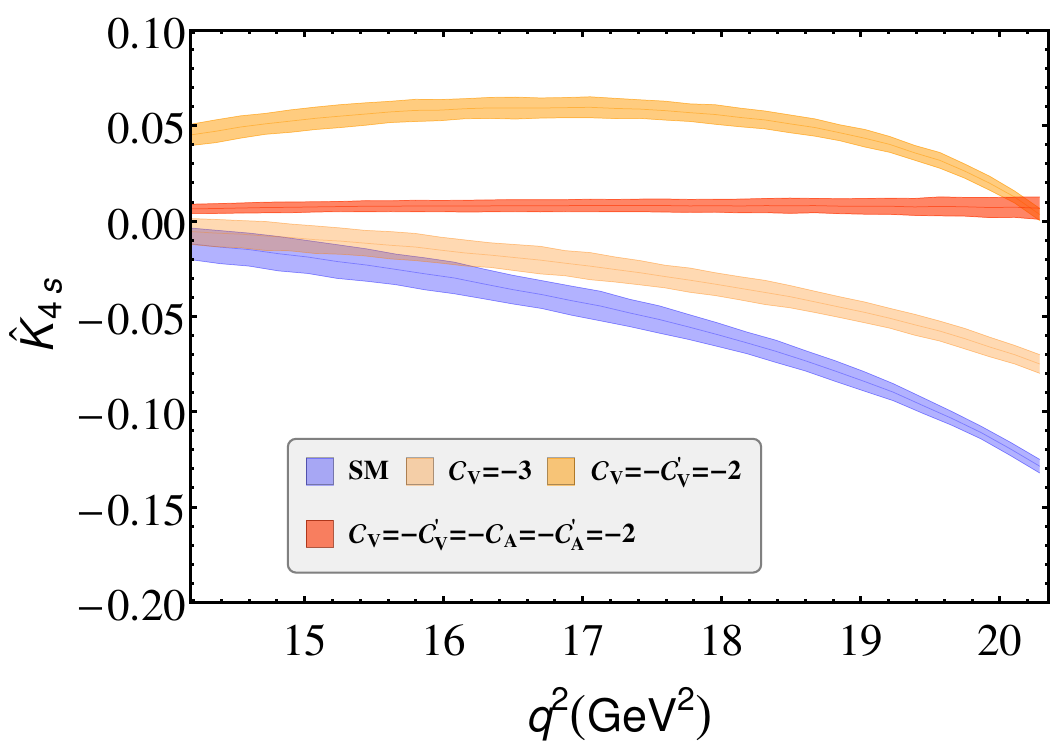}
		\includegraphics[scale=0.4]{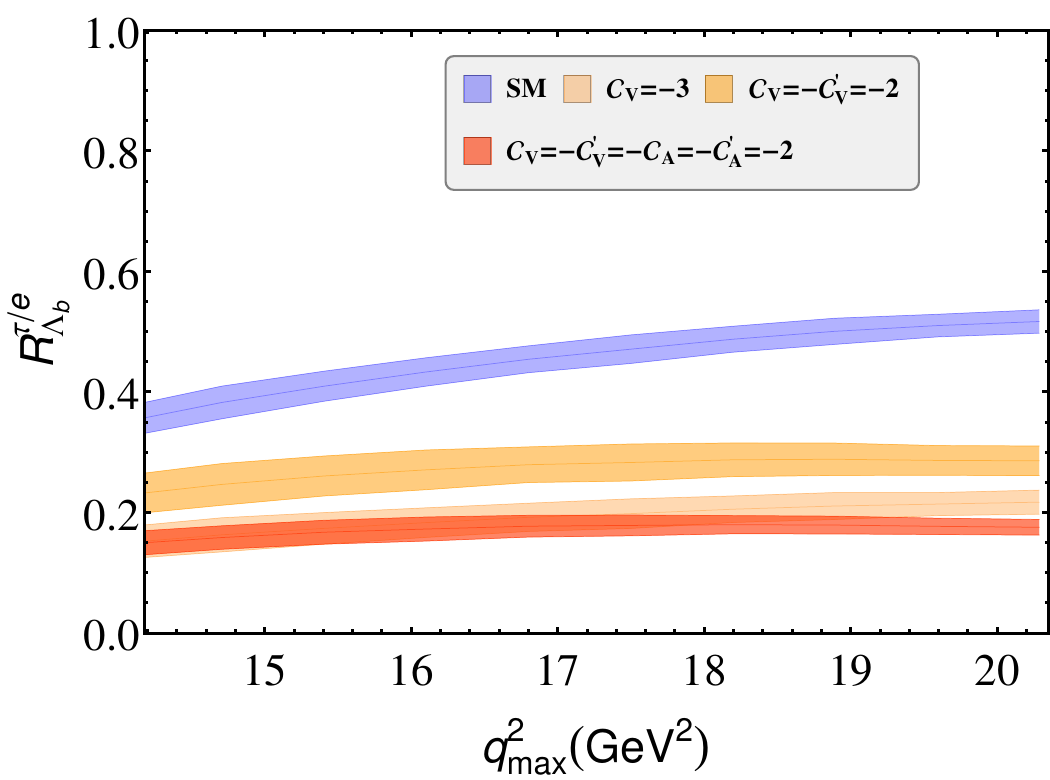}
		\caption{The $\Lambda_b\to\Lambda(\to p\pi)\tau^+\tau^-$ observables in the presence of VA couplings. The bands correspond to the uncertainties discussed in text. Since the OPE at large $q^2$ does not capture local resonance structures, the distributions can be locally off from the OPE predictions.\label{fig:VA}}
	\end{center}
\end{figure}
\begin{figure}[h!]
	\begin{center}
		\includegraphics[scale=0.4]{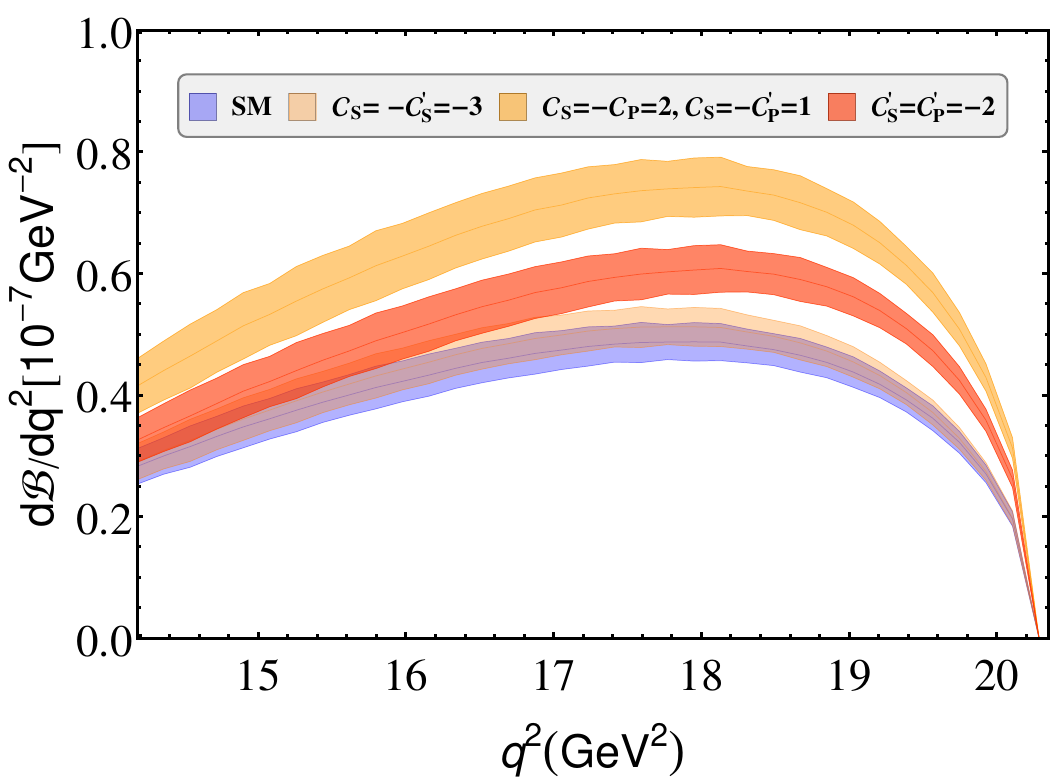}
		\includegraphics[scale=0.4]{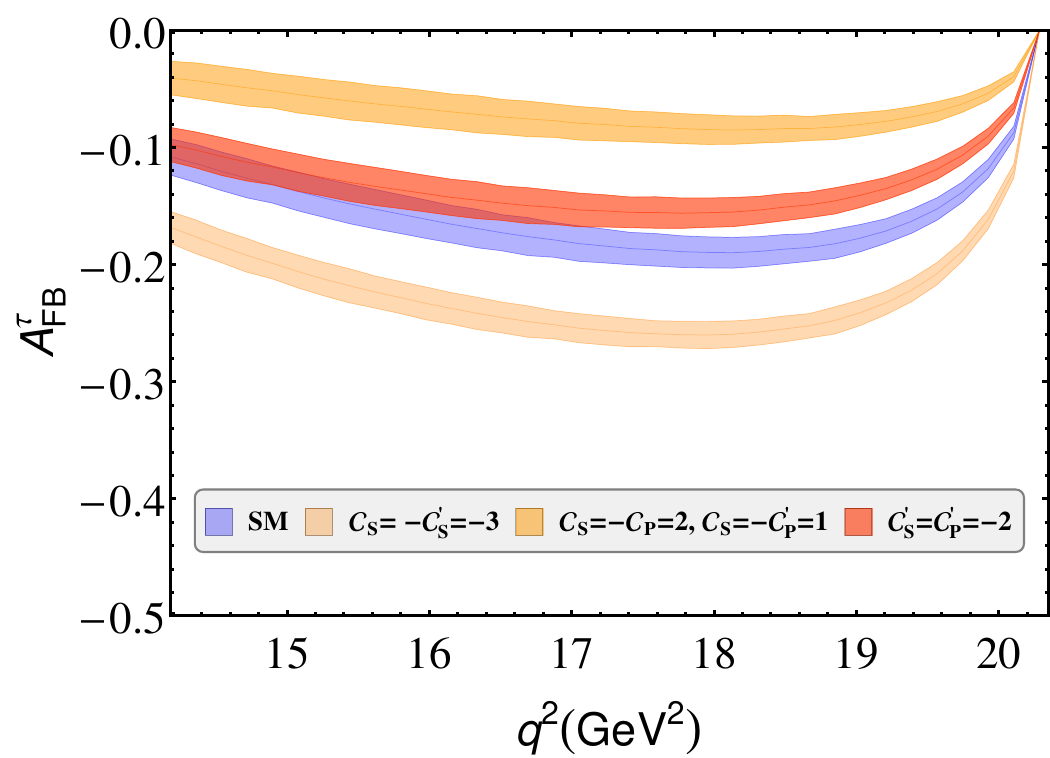}
		\includegraphics[scale=0.4]{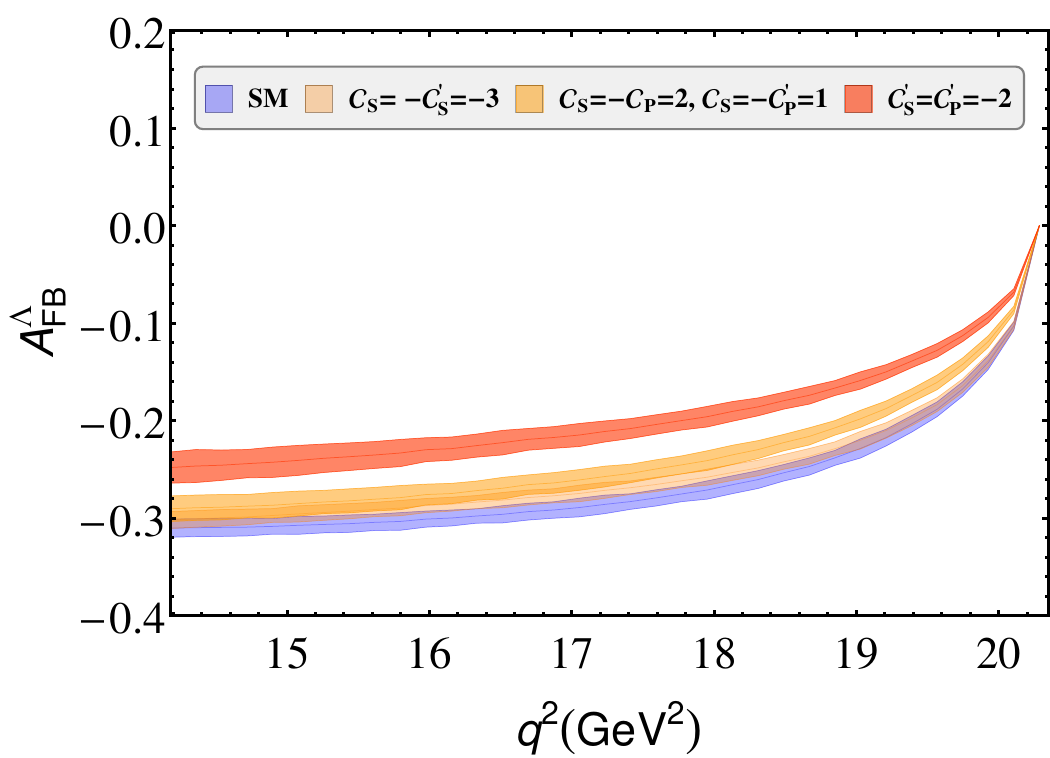}
		\includegraphics[scale=0.4]{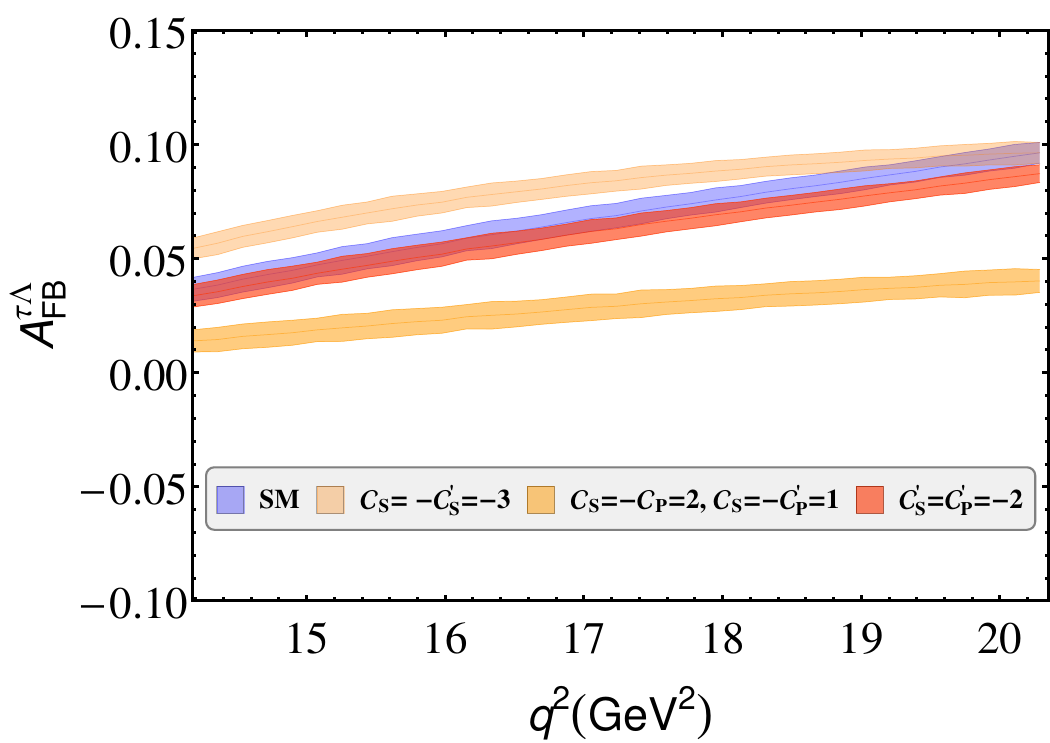}
		\includegraphics[scale=0.41]{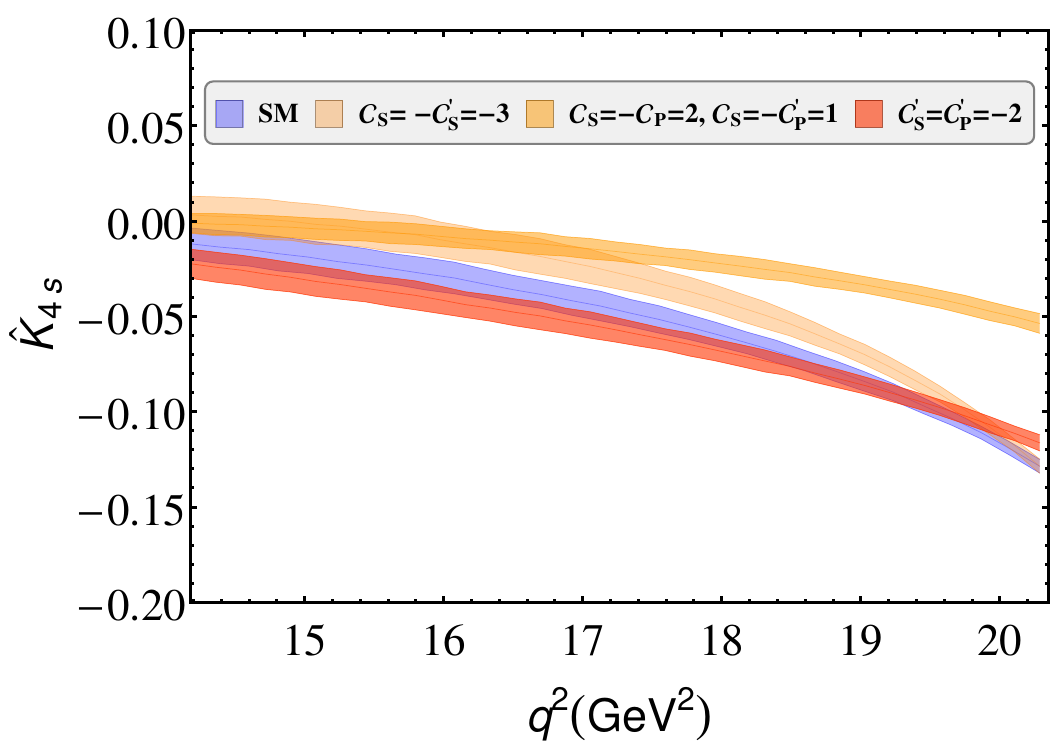}
		\includegraphics[scale=0.4]{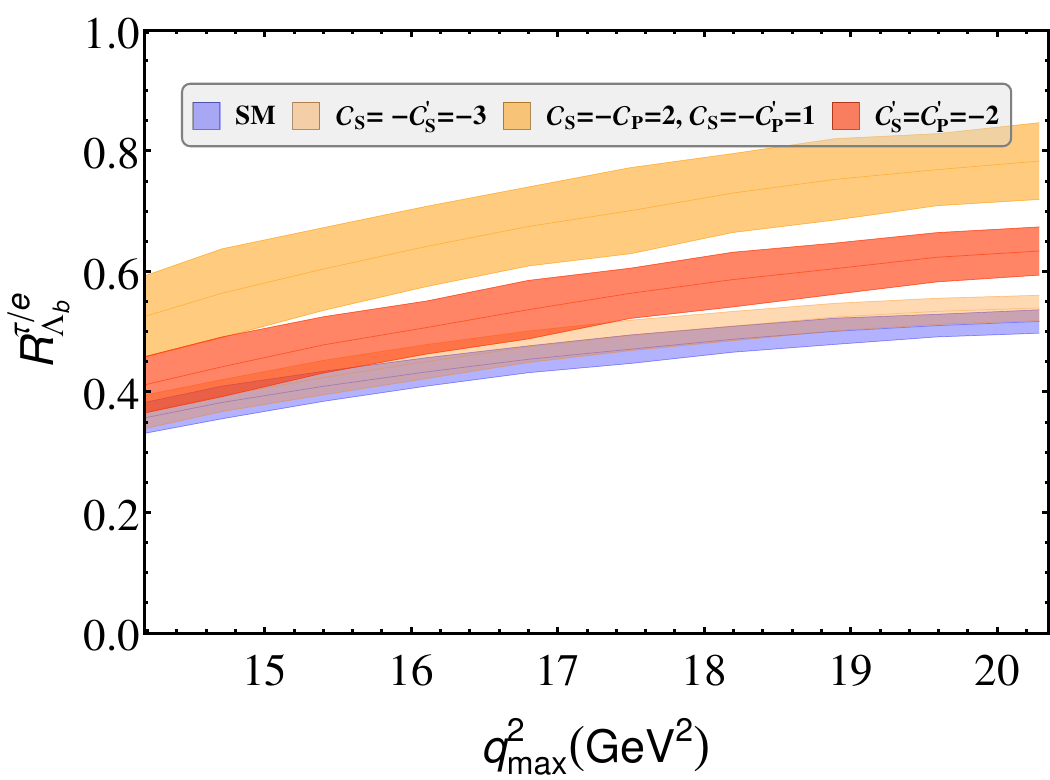}
		\caption{The $\Lambda_b\to\Lambda(\to p\pi)\tau^+\tau^-$ observables in the presence of SP couplings. The bands correspond to the uncertainties discussed in text. Since the OPE at large $q^2$ does not capture local resonance structures, the distributions can be locally off from the OPE predictions.\label{fig:SP}}
	\end{center}
\end{figure}
%
We now discuss the observables in the presence of the NP couplings. The VA and the SP couplings will be considered separately. Due to nonavailability of any data on $b\to s\tau^+\tau^-$ transition, currently the new physics couplings are very poorly constrained. For VA couplings, following \cite{Kamenik:2017ghi} we consider the following three scenarios
\begin{eqnarray}\label{eq:VAWilson}
	\mC_V &=& -3\, \nn\\
	\mC_V &=& - \mC_V^{\prime} = -2\, \\
	\mC_V &=& -\mC_V^\prime = -\mC_A = -\mC_A^\prime = -2\, .\nn
\end{eqnarray}
These ranges of couplings are consistent with the existing direct bounds on $B\to K^+\tau^+\tau^-$ \cite{Bobeth:2011st} and $B_s\to \tau^+\tau^-$ \cite{Aaij:2017xqt}, and the bounds on $B\to K^\ast\nu\bar{\nu}$ due to heavy NP respecting SM $SU(2)_L$ gauge invariance \cite{Alonso:2015sja}.

In Fig.~\ref{fig:VA} we present the effects of $\mC_{V,A}^{(\prime)}$ couplings. The three cases in (\ref{eq:VAWilson}) are separately considered and in each case the rest of the couplings are SM like. We find that the branching ratio gets reduced for all the three scenarios. The leptonic forward-backward asymmetry can be both greater or smaller than the SM value. Note that in the absence of chirality flipped operators, \emph{i.e.,} when $\mC_{V,A}^{\prime}=0$, the hadronic forward-backward asymmetry is independent of short-distance Wilson coefficients due to the low recoil symmetry of the form factors \cite{Boer:2014kda}. This is also the reason why in the scenario $\mC_V= -3$, the $A^\Lambda_{\rm FB}$ is SM like as can be seen from Fig.~\ref{fig:VA}. In the last scenario of (\ref{eq:VAWilson}), $\mC_V = -\mC_V^\prime = -\mC_A = -\mC_A^\prime = -2$ the asymmetry $A^{\Lambda}_{\rm FB}$ has opposite sign than SM. For this scenario we notice sign flip in $A^{\tau\Lambda}_{\rm FB}$ and $\hat{K}_{4sc}$ also. For all the three cases, the value of $R^{\tau/e}_{\Lambda_b}$ is smaller than the SM value.

For the scalar couplings, indirect bounds from $b\to s\gamma$ and $b\to s\ell^+\ell^-$, and direct bounds from the limits on $B_s\to \tau^+\tau^-$ branching ratio, the exclusive semi-leptonic mode $B^+\to K^+\tau^+\tau^-$ and its inclusive counterpart $B\to X_s\tau^+\tau^-$ were studied in \cite{Bobeth:2011st}. The obtained bounds are  $|\mC_S \mp \mC_P| < 0.4\frac{2\pi}{\alpha_e}\, $, $ |\mC_S^\prime \mp \mC_P^\prime| < 0.4\frac{2\pi}{\alpha_e}\,$ which are extremely weak. Therefore, for illustrative purpose we choose three sets of couplings (i) $\mC_S=-\mC_S^\prime=-3$, (ii) $\mC_S=-\mC_P=2, \mC_S=-\mC_P^\prime=-1$, and (iii) $\mC_S^\prime=\mC_P^\prime=-2$. In Fig.~\ref{fig:SP} we show the plots for these three sets of couplings. The branching ratio and hence $R^{\tau/e}_{\Lambda}$ are enhanced for these couplings. However, for these set of couplings we did not notice any sign flip for the asymmetries $A^\Lambda_{\rm FB}$ and $A^{\tau\Lambda}_{\rm FB}$.

\section{Summary \label{sec:summary}}
In this paper we have presented a full angular distribution of $\Lambda_b\to\Lambda(\to N\pi)\ell^+\ell^-$ decay for heavy leptons and unpolarized $\Lambda_b$ in an operator basis that includes the SM operators, new vector and axial-vector operators, and scalar and pseudo-scalar operators. We have worked in the transversity basis and expressed the angular coefficients in terms of the transversity amplitudes. The $\Lambda_b\to \Lambda$ hadronic matrix elements are parametrized in terms of helicity form factors whose $q^2$ dependence are known from fit to calculations in lattice QCD. For our numerical analysis we have studied the mode $\Lambda_b\to\Lambda(\to p\pi)\tau^+\tau^-$. We have presented the SM determinations of several observables including the lepton flavor universality sensitive observable $R^{\tau/e}_{\Lambda_b}$. We have also explored the effects of the NP couplings on the observables.

\section*{Acknowledgements}
The author is supported by the DST, Govt. of India under INSPIRE Faculty Fellowship (award letter number DST-INSPIRE/04/2016/002620).

\appendix

\section{SM Wilson coefficients \label{app:C910}}
The standard model effective Wilson coefficients $\mC_{7,9}^{\rm eff}$ are given as \cite{Grinstein:2004vb}
\begin{eqnarray}
\mC_9^{\rm eff} &=& \mathcal{C}_9^{\rm SM} + h(0,q^2) \bigg[ \frac{4}{3}\mC_1 + \mC_2 + \frac{11}{2}\mC_3 - \frac{2}{3}\mC_4 + 52\mC_5 - \frac{32}{3}\mC_6 \bigg]\, \nn\\ &-& \frac{1}{2}h(m_b,q^2) \bigg[7\mC_3 + \frac{4}{3}\mC_4 + 76\mC_5 + \frac{64}{3}\mC_6 \bigg] + \frac{4}{3} \bigg[\mC_3 + \frac{16}{3}\mC_5 + \frac{16}{9}\mC_6 \bigg]\, \nn\\ &+& 8\frac{m_c^2}{q^2} \bigg[ \frac{4}{9}\mC_1 + \frac{1}{3}\mC_2 + 2\mC_3 + 20\mC_5 \bigg]\, \nn\\ &+& \frac{\alpha_s}{4\pi} \bigg[\mC_1(B(q^2)+4C(q^2)) - 3\mC_2(2B(q^2)-C(q^2)) - \mC_8 F_8^{(9)}(q^2) \bigg]\, ,\\
\mC_7^{\rm eff} &=& \mC_7^{\rm SM} - \frac{1}{3}\bigg[\mC_3 + \frac{4}{3}\mC_4 + 20\mC_5 + \frac{80}{3}\mC_6 \bigg] + \frac{\alpha_s}{4\pi} \bigg[ (\mC_1 - 6\mC_2)A(q^2) - \mC_8 F_8^{(7)}(q^2) \bigg]\, ,
\end{eqnarray}
where $\mathcal{C}_{7,9}^{\rm SM}$ are the SM values at the scale of $b$-quark mass \cite{Altmannshofer:2008dz}, $\mu = m_b=4.8$ GeV. Rest of the Wilson coefficients are also taken from \cite{Altmannshofer:2008dz}. The charm and the $b$-quark loop functions read \cite{Buras:1994dj,Misiak:1992bc}
\begin{eqnarray}
h(m_q, q^2) &=& -\frac{8}{9}\ln\frac{m_q}{m_b} + \frac{8}{27} + \frac{4}{9}x - \frac{2}{9}(2+x)|1-x|^{1/2} \left\{
\begin{array}{ll}
\ln\left| \frac{\sqrt{1-x} + 1}{\sqrt{1-x} - 1}\right| - i\pi, & x \equiv \frac{4 m_c^2}{ q^2} < 1,  \\ & \\ 2 \arctan \frac{1}{\sqrt{x-1}}, & x \equiv \frac {4 m_c^2}{ q^2} > 1,
\end{array}
\right. \\
h(0,q^2) & =& \frac{8}{27} - \frac{4}{9} \ln\frac{q^2}{m_b^2}  + \frac{4}{9} i\pi.
\end{eqnarray}
The loop functions $A(q^2), B(q^2), C(q^2)$ and $F_8^{(7,9)}$ are taken from \cite{Beneke:2001at,Seidel:2004jh}.

\section{Transversity amplitudes \label{sec:TAs2}}
The $\Lambda_b \to \Lambda$ hadronic matrix elements for different operators can be parametrized in terms of ten helicity form factors $f^V_{t,0,\perp}$, $f^A_{t,0,\perp}$, $f^T_{0,\perp}$, $f^{T5}_{0,\perp}$ \cite{Feldmann:2011xf} and spinor matrix elements. For the operators (\ref{eq:opbasis}), the hadronic matrix elements parametrization in terms of the spinor bilinears are summarized in Appendix~\ref{sec:hme}. The spinor bilinears can be found in the Appendix of \cite{Das:2018sms}. The helicity amplitudes are defined as the projection of the hadronic matrix elements on to the direction of the polarization of the virtual gauge boson that decays to a dilepton pair. For VA operators the detailed derivations of the helicity amplitudes for massless leptons can be found in \cite{Das:2018sms}. For completeness we write down their expressions
\begin{eqnarray}
H^{L(R),\plpl}_{\rm VA,0} &=& f^V_0(\mLb + \mL) \sqrt{\frac{s_-}{q^2}} \mC^{L,(R)}_{\rm VA,+} - f^A_0(\mLb - \mL) \sqrt{\frac{s_+}{q^2}} \mC^{L,(R)}_{\rm VA,-} \, \nn\\ &+& \frac{2m_b}{q^2} \bigg( f^T_0 \sqrt{q^2 s_-} - f^{T5}_0 \sqrt{q^2 s_+} \bigg)  \mC_7^{\rm eff}\, ,\\
H^{L(R),\mimi}_{\rm VA,0} &=& f^V_0(\mLb + \mL) \sqrt{\frac{s_-}{q^2}} \mC^{L,(R)}_{\rm VA,+} + f^A_0(\mLb - \mL) \sqrt{\frac{s_+}{q^2}} \mC^{L,(R)}_{\rm VA,-} \, \nn\\ &+& \frac{2m_b}{q^2} \bigg( f^T_0 \sqrt{q^2 s_-} + f^{T5}_0 \sqrt{q^2 s_+} \bigg)  \mC_7^{\rm eff}\, ,\\
H^{L(R),\mipl}_{\rm VA,+} &=& -f^V_\perp \sqrt{2s_-} \mC^{L,(R)}_{\rm VA,+} + f^A_\perp \sqrt{2s_+} \mC^{L,(R)}_{\rm VA,-} \, \nn\\ &-& \frac{2m_b}{q^2} \bigg( f^T_\perp(\mLb + \mL) \sqrt{2s_-} - f^{T5}_\perp (\mLb - \mL) \sqrt{2s_+}   \bigg)\mC_7^{\rm eff}\,,~~~\\
H^{L(R),\plmi}_{\rm VA,-} &=& -f^V_\perp \sqrt{2s_-} \mC^{L,(R)}_{\rm VA,+} - f^A_\perp \sqrt{2s_+} \mC^{L,(R)}_{\rm VA,-} \, \nn\\ &-& \frac{2m_b}{q^2} \bigg( f^T_\perp(\mLb + \mL) \sqrt{2s_-} + f^{T5}_\perp (\mLb - \mL) \sqrt{2s_+}   \bigg)\mC_7^{\rm eff}\,.
\end{eqnarray}

If the leptons are massive, then there are additional timelike helicity amplitudes that can be defined. The timelike polarization of the gauge boson $\bar{\epsilon}^\mu=q^\mu/\sqrt{q^2}$ leads to the following conservation laws
\begin{equation}
	q^\mu(\bar{\ell}\gamma_\mu\ell)=0 \, ,\quad 	q^\mu(\bar{\ell}\gamma_\mu\gamma_5\ell) = 2mi(\bar{\ell}\gamma_5\ell)\, .
\end{equation} 
These equations imply that the timelike components of the gauge boson can only couple to the axial-vector current and therefore can not have separate left- and right-handed parts. The two non-vanishing helicity amplitudes that we find are
\begin{eqnarray}
	H^{\plpl}_t &=& 2 \bigg( f^A_t (\mLb + \mL) \sqrt{\frac{s_-}{q^2}} ( \mC_{10} + \mC_A - \mC_A^\prime) \,\nn\\
	&-& f^V_t (\mLb - \mL) \sqrt{\frac{s_+}{q^2}} ( \mC_{10} + \mC_A + \mC_A^\prime)   \bigg)\, ,\\
	H^{\mimi}_t &=& -2 \bigg( f^A_t (\mLb + \mL) \sqrt{\frac{s_-}{q^2}} ( \mC_{10} + \mC_A - \mC_A^\prime) \,\nn\\
	&+& f^V_t (\mLb - \mL) \sqrt{\frac{s_+}{q^2}} ( \mC_{10} + \mC_A + \mC_A^\prime)   \bigg)\, .
\end{eqnarray}
From these helicity amplitudes, we construct the following transversity amplitudes corresponding to the VA currents
\begin{align}\label{eq:TAVA1}
& A_{\perp_0}^{L(R)} = \frac{N}{\sqrt{2}}\bigg[H^{L(R),\plpl}_0 + H^{L(R),\mimi}_0  \bigg]\, ,\\
& A_{\|_0}^{L(R)} = \frac{N}{\sqrt{2}}\bigg[H^{L(R),\plpl}_0 - H^{L(R),\mimi}_0  \bigg]\, ,\\
& A_{\perp_1}^{L(R)} = \frac{N}{\sqrt{2}} \bigg[H^{L(R),\mipl}_+ + H^{L(R),\plmi}_- \bigg]\, ,\\
& A_{\|_1}^{L(R)} = \frac{N}{\sqrt{2}} \bigg[H^{L(R),\mipl}_+ - H^{L(R),\plmi}_- \bigg]\, ,\\
& A_{\perp_t} = \frac{N}{\sqrt{2}} \bigg[ H^{\plpl}_t + H^{\mimi}_t \bigg]\, ,\\ 
\label{eq:TAVA2}
& A_{\|_t} = \frac{N}{\sqrt{2}} \bigg[ H^{\plpl}_t - H^{\mimi}_t \bigg]\, ,
\end{align} 

The helicty amplitudes corresponding to the scalar operators are derived in details in Ref.~\cite{Das:2018sms}. Here we define the transversity amplitudes in terms of them
\begin{align}\label{eq:TAscalar1}
& A_{\rm S\perp} =  \frac{N}{\sqrt{2}}\bigg[ H^{L,\plpl}_{\rm SP} + H^{R,\plpl}_{\rm SP} + H^{L,\mimi}_{\rm SP} + H^{R,\mimi}_{\rm SP} \bigg]\, ,\\
& A_{\rm S \|} =  \frac{N}{\sqrt{2}}\bigg[ H^{L,\plpl}_{\rm SP} + H^{R,\plpl}_{\rm SP} - H^{L,\mimi}_{\rm SP} - H^{R,\mimi}_{\rm SP}  \bigg]\, ,\\
& A_{\rm P\perp} =  \frac{N}{\sqrt{2}}\bigg[ H^{L,\plpl}_{\rm SP} - H^{R,\plpl}_{\rm SP} + H^{L,\mimi}_{\rm SP} - H^{R,\mimi}_{\rm SP}  \bigg]\, ,\\
\label{eq:TAscalar2}
& A_{\rm P \|} =  \frac{N}{\sqrt{2}}\bigg[ H^{L,\plpl}_{\rm SP} - H^{R,\plpl}_{\rm SP} - H^{L,\mimi}_{\rm SP} + H^{R,\mimi}_{\rm SP}  \bigg]\, .
\end{align}

\section{Angular coefficients \label{sec:angular}}
From Sec.~\ref{sec:angdist} we recall that the angular coefficients are written as
\begin{equation}
K_{\{\cdots\}} = \mathcal{K}_{\{\cdots\}} + \frac{m_\ell}{\sqrt{q^2}} \mathcal{K}_{\{\cdots\}}^\prime + \frac{m_\ell^2}{q^2}\mathcal{K}_{\{\cdots\}}^{\prime\prime}\, ,
\end{equation}
where $\{\cdots\}$ correspond to the suffixes $1ss, 1cc, 1c, 2ss,2cc,2c,3sc,3s,4sc,4s$. In terms of the transversity amplitudes the expressions of $\mathcal{K}, \mathcal{K}^\prime \mathcal{K}^{\prime\prime}$ read as
\begin{eqnarray}\label{eq:KKprime1}
\mK_{1ss} &=& \frac{1}{4} \bigg( 2|\ARpa0|^2 + |\ARpa1|^2 + 2|\ARpe0|^2 + |\ARpe1|^2 + \{ R \leftrightarrow L  \} \bigg) \nn\\&+& \frac{1}{4}\bigg( |A_{\rm S\perp}|^2 + |A_{\rm P\perp}|^2 + \{ \perp \leftrightarrow \| \} \bigg)\, ,~~~~\\
\mK_{1ss}^\prime &=& \re\bigg( A_{\|t}A_{\rm P\|}^\ast + A_{\perp t} A^\ast_{\rm P\perp}  \bigg)\, , \\
\mK_{1ss}^{\prime\prime} &=& -\bigg( |A^R_{\|_0}|^2 + |A^R_{\perp_0}|^2 + \{ R \leftrightarrow L \} \bigg) + \bigg( |A_{\perp t}|^2 - |A_{\rm S\perp}|  + \{ \perp \leftrightarrow \| \}\bigg)  \nn\\ &+& 2\re\bigg( A^R_{\perp_0}A^{\ast L}_{\perp_0} + A^R_{\perp_1}A^{\ast L}_{\perp_1} + \{ \perp \leftrightarrow \| \}  \bigg)\, , \\
\mK_{1cc} &=& \frac{1}{2}\bigg( |\ARpa1|^2 + |\ARpe1|^2 + \{R \leftrightarrow L \} \bigg) + \frac{1}{4}\bigg( |A_{\rm P\perp}|^2 + |A_{\rm S\perp}|^2 + \{\perp \leftrightarrow \| \} \bigg) \, ,\\
\mK_{1cc}^\prime &=& \re\bigg( A_{\| t}A^{\ast}_{\rm P\|} + A_{\perp t}A^{\ast}_{\rm P\perp}   \bigg) ,\\
\mK_{1cc}^{\prime\prime} &=& \bigg( |A^R_{\|_0}|^2 - |A^R_{\|_1}|^2 + |A^R_{\perp_0}|^2 - |A^R_{\perp_1}|^2 + \{ R \leftrightarrow L \} \bigg) + \bigg( |A_{\perp t}|^2 - |A_{\rm S\perp}|^2 + \{ \perp \leftrightarrow \| \}\bigg)\, \nn\\ &+& 2\re\bigg( A^R_{\perp_0}A^{\ast L}_{\perp_0} + A^R_{\perp_1}A^{\ast L}_{\perp_1} + \{\perp \leftrightarrow \| \} \bigg)\, , \\
\mK_{1c} &=& -\beta_\ell \bigg( A^R_{\perp_1}A^{\ast R}_{\|_1} - \{ R \leftrightarrow L \}  \bigg)\, ,\\
\mK_{1c}^\prime &=& \beta_\ell \re\bigg( A_{\rm S\perp}A^{\ast R}_{\perp_0} + A_{\rm S\perp}A^{\ast L}_{\perp_0} + \{ \perp \leftrightarrow \| \} \bigg)\, ,\\
\mK_{1c}^{\prime\prime} &=& 0\, ,\\
\mK_{2ss} &=& \frac{\alpha_\Lambda}{2} \re\bigg( 2 A^R_{\perp_0}A^{\ast R}_{\|_0} + A^R_{\perp_1}A^{\ast R}_{\|_1} + \{ R \leftrightarrow L \} \bigg) + \frac{\alpha_\Lambda}{2}\re\bigg( A_{\rm P\perp}A^{\ast}_{\rm P\|} + A_{\rm S\perp}A^{\ast}_{\rm S\|} \bigg) \, ,\\
\mK_{2ss}^\prime &=& \alpha_\Lambda \re\bigg( A_{\perp t}A^{\ast}_{\rm P\|} + A_{\| t}A^{\ast}_{\rm P\perp} \bigg)\, ,\\
\mK_{2ss}^{\prime\prime} &=& -2\alpha_\Lambda \re\bigg( A^R_{\perp_0}A^{\ast R}_{\|_0} + A^L_{\perp_0}A^{\ast L}_{\|_0} - A^R_{\perp_0}A^{\ast L}_{\|_0} - A^R_{\|_0}A^{\ast L}_{\perp_0} - A^R_{\perp_1}A^{\ast L}_{\|_1} - A^R_{\|_1}A^{\ast L}_{\perp_1} \nn\\&-& A_{\perp t}A^\ast_{\|t} + A_{\rm S\perp}A^\ast_{\rm S\|}  \bigg)\, ,\\
\mK_{2cc} &=& \alpha_\Lambda \re \bigg( A^R_{\perp_1}A^{\ast R}_{\|_1} + A^L_{\perp_1}A^{\ast L}_{\|_1} \bigg) + \frac{\alpha_\Lambda}{2}\re\bigg( A_{\rm P\perp}A^\ast_{\rm P\|} + A_{\rm S\perp}A^\ast_{\rm S\|}\bigg) \, ,\\
\mK_{2cc}^\prime &=& \alpha_\Lambda \re\bigg( A_{\perp t}A^{\ast}_{\rm P\|} +A_{\| t}A^{\ast}_{\rm P\perp} \bigg)\, ,\\
\mK_{2cc}^{\prime\prime} &=& -2\alpha_\Lambda \re\bigg( A^R_{\perp_1}A^{\ast R}_{\|_1} - A^R_{\perp_0}A^{\ast R}_{\|_0} + A^L_{\perp_1}A^{\ast L}_{\|_1} - A^L_{\perp_0}A^{\ast L}_{\|_0} - A^R_{\perp_0}A^{\ast L}_{\|_0} - A^R_{\|_0}A^{\ast L}_{\perp_0}\nn\\ &-& A^R_{\perp_1}A^{\ast L}_{\|_1} - A^R_{\|_1}A^{\ast L}_{\perp_1} - A_{\perp t}A^{\ast}_{\|t} + A_{\rm S\perp}A^{\ast}_{\rm S\|}  \bigg)\, ,\\
\mK_{2c} &=& -\frac{\alpha_\Lambda\beta_\ell}{2} \re\bigg( |\ARpe1|^2 + |\ARpa1|^2 - \{ R \leftrightarrow L \}   \bigg)\, ,\\
\mK_{2c}^\prime &=& \alpha_\Lambda\beta_\ell \re\bigg( A_{\rm S\perp}A^{\ast R}_{\|_0} + A_{\rm S\|}A^{\ast R}_{\perp_0} + \{ R \leftrightarrow L \} \bigg)\, ,\\
\mK_{2c}^{\prime\prime} &=& 0\, ,\\
\mK_{3sc} &=& \frac{\alpha_\Lambda}{\sqrt{2}} \im\bigg( \ARpe1\AsRpe0 - \ARpa1\AsRpa0 + \{ R \leftrightarrow L \} \bigg)\, ,\\
\mK_{3sc}^\prime &=& 0\, ,\\
\mK_{3sc}^{\prime\prime} &=& 2\sqrt{2}\alpha_\Lambda \im\bigg( A^R_{\|_1}A^{\ast R}_{\|_0} - A^R_{\perp_1}A^{\ast R}_{\perp_0} + \{ R \leftrightarrow L \}  \bigg)\, ,\\
\mK_{3s} &=& \frac{\alpha_\Lambda\beta_\ell}{\sqrt{2}} \im\bigg( A^R_{\perp_1}A^{\ast R}_{\|_0} - A^R_{\|_1}A^{\ast R}_{\perp_0} - \{ R \leftrightarrow L \} \bigg)\, ,\\
\mK_{3s}^\prime &=& \frac{\alpha_\Lambda\beta_\ell}{\sqrt{2}} \im\bigg( A^R_{\|_1}A^\ast_{\rm S\|} - A^R_{\perp_1}A^\ast_{\rm S\perp} + \{ R \leftrightarrow L \}  \bigg)\, ,\\
\mK_{3s}^{\prime\prime} &=& 0\, ,\\
\mK_{4sc} &=& \frac{\alpha_\Lambda}{\sqrt{2}} \re\bigg( A^R_{\perp_1}A^{\ast R}_{\|_0} - A^R_{\|_1}A^{\ast R}_{\perp_0} + \{ R \leftrightarrow L \} \bigg)\, ,\\
\mK_{4sc}^\prime &=& 0\, ,\\
\mK_{4sc}^{\prime\prime} &=& 2\sqrt{2}\alpha_\Lambda \re\bigg( A^R_{\|_1}A^{\ast R}_{\perp_0} - A^R_{\perp_1}A^{\ast R}_{\|_0} + \{ R \leftrightarrow L \}  \bigg)\, ,\\
\mK_{4s} &=& \frac{\alpha_\Lambda\beta_\ell}{\sqrt{2}} \re \bigg( A^R_{\perp_1}A^{\ast R}_{\perp_0} - A^R_{\|_1}A^{\ast R}_{\|_0} - \{ R \leftrightarrow L \} \bigg)\, ,\\
\mK_{4s}^\prime &=& \frac{\alpha_\Lambda\beta_\ell}{\sqrt{2}} \re \bigg( A^R_{\|_1}A^\ast_{\rm S\perp} - A^R_{\perp_1}A^\ast_{\rm S\|} + A_{\rm S\perp}A^{\ast R}_{\|_1} - A_{\rm S\|}A^{\ast R}_{\perp_1} + \{ R \leftrightarrow L \bigg)\, ,\\
\label{eq:KKprime2}
\mK_{4s}^{\prime\prime} &=& 0\, .
\end{eqnarray}
For our SM results, we agree\footnote{The normalization of our time-like transversity amplitudes differ from Ref.\cite{Blake:2017une} by a factor of $\sqrt{2}$.} with \cite{Blake:2017une}

\section{$\Lambda_b \to \Lambda$ hadronic matrix elements \label{sec:hme}}
A convenient choice for the parametrization of the $\Lambda_b \to \Lambda$ hadronic matrix elements is the so called helicty basis \cite{Feldmann:2011xf} in terms of which the matrix elements for the vector current is
\begin{align}\label{eq:VAhme1}
\langle \Lambda(k,s_k)|\bar{s}\gamma^\mu b |\Lambda(p,s_p)\rangle =& \bar{u}(k,s_k)\Bigg[f^V_t(q^2)(\mLb-\mL)\frac{q^\mu}{q^2}\nn\\ +& f^V_0(q^2) \frac{\mLb+\mL}{s_+} \{p^\mu + k^\mu  - \frac{q^\mu}{q^2}(\mmLb - \mmL) \} \nn\\ + & f^V_\perp(q^2) \{ \gamma^\mu - \frac{2\mL}{s_+}p^\mu - \frac{2\mLb}{s_+}k^\mu \} \Bigg]u(p,s_p)\, ,
\end{align}
and the axial-vector current is
\begin{align}\label{eq:VAhme2}
\langle \Lambda(k,s_k)|\bar{s}\gamma^\mu\gamma_5 b |\Lambda(p,s_p)\rangle =& - \bar{u}(k,s_k) \gamma_5 \Bigg[ f_t^A(q^2) (\mLb + \mL) \frac{q^\mu}{q^2} \nn\\ +& f_0^A(q^2) \frac{\mLb - \mL}{s_-} \{p^\mu + k^\mu - \frac{q^\mu}{q^2} (\mmLb - \mmL) \} \nn\\ + & f_\perp^A(q^2) \{\gamma^\mu + \frac{2\mL}{s_-}p^\mu - \frac{2\mLb}{s_-}k^\mu \}  \Bigg] u(p,s_p)\, .
\end{align}
The matrix elements for the scalar and the pseudo-scalar currents are
\begin{align}
\langle \Lambda(k,s_k)|\bar{s} b |\Lambda(p,s_p)\rangle =& f^V_t(q^2) \frac{\mLb-\mL}{m_b} \bar{u}(k,s_k)u(p,s_p)\, ,\\
\langle \Lambda(k,s_k)|\bar{s} \gamma_5 b |\Lambda(p,s_p)\rangle =& f_t^A(q^2) \frac{\mLb + \mL}{m_b} \bar{u}(k,s_k) \gamma_5 u(p,s_p)\, ,
\end{align}
where we have neglected the mass of the strange quark.
For the dipole operators we get 
\begin{eqnarray}
\langle \Lambda |\bar{s}i q_\nu\sigma^{\mu\nu}b|\Lambda_b\rangle &=& -\bar{u}(k,s_k)\Bigg[ f^T_0(q^2) \frac{q^2}{s_+}\Bigg(p^\mu + k^\mu - \frac{q^\mu}{q^2}(\mmLb - \mmL) \Bigg) \,\nn\\ &+& f^T_\perp (\mLb+\mL) \Bigg( \gamma^\mu - \frac{2\mL}{s_+}p^\mu - \frac{2\mLb}{s_+}k^\mu \Bigg) \Bigg]u(p,s_p)\, ,
\end{eqnarray}
and 
\begin{eqnarray}
\langle \Lambda |\bar{s}i q_\nu\sigma^{\mu\nu}\gamma_5 b|\Lambda_b\rangle &=& -\bar{u}(k,s_k)\gamma_5 \Bigg[f^{T5}_0 \frac{q^2}{s_-} \Bigg( p^\mu + k^\mu - \frac{q^\mu}{q^2}(\mmLb - \mmL) \Bigg) \, \nn\\ &+& f^{T5}_\perp (\mLb - \mL) \Bigg( \gamma^\mu + \frac{2\mL}{s_-}p^\mu - \frac{2\mLb}{s_-}k^\mu \Bigg) \Bigg] u(p,s_p)\, .
\end{eqnarray}
%

\section{Numerical inputs \label{sec:inputs}}

In the following table we collect the numerical values of the inputs used in the paper.

\begin{longtable}{cr|cr}
	\hline\hline
	inputs  & values  & inputs  & values   \\ 
	\hline 
	$ \alpha_e(m_b) $  & $1/127.925(16)$ \cite{Patrignani:2016xqp} & $|V_{tb}V_{ts}^\ast|$  & $0.0401 \pm 0.0010$ \cite{Bona:2006ah}  \\ 
	$m_c(\overline{\text{MS}})$ & 1.28 GeV \cite{Patrignani:2016xqp} & $\mLb$ & 5.619 GeV \cite{Patrignani:2016xqp} \\ 
	$\mu_b$ & $4.8$ GeV \cite{Altmannshofer:2008dz} & $\mL$ & $1.115$ GeV \cite{Patrignani:2016xqp} \\
	$m_b(\overline{\text{MS}})$ & $4.2$ GeV \cite{Altmannshofer:2008dz} & $\tau_{\Lambda_b}$ & $(1.470\pm 0.010)\times 10^{-12} s$ \cite{Patrignani:2016xqp} \\  
	$m_{b}({\rm pole})$ & $4.8$ GeV \cite{Altmannshofer:2008dz}  & $m_{B^0}$ & $5.279$ GeV \cite{Detmold:2016pkz} \\ 
	$\alpha_s(m_b)$ & $0.214$ \cite{Altmannshofer:2008dz}  & $m_K$ & $0.494$ GeV \cite{Detmold:2016pkz} \\ 
	$|V_{cb}|$ & $(4.15 \pm 0.07)\times 10^{-2}$ \cite{Bona:2006ah} & $\alpha_\Lambda$ & $0.642\pm0.013$ \cite{Patrignani:2016xqp}\\
	$\alpha_s(M_Z)$ & $0.1181\pm 0.0011$ \cite{Patrignani:2016xqp} & $m_c(m_c)$ & $1.28\pm 0.03$ GeV \cite{Patrignani:2016xqp}\\
	\hline\caption{List of inputs and their values. \label{tab:inputs}}
\end{longtable}

\end{document}